%% file: SN2023.tex
\documentclass[twocolumn,twocolappendix]{aastex631}
\usepackage{CJK}
\usepackage[utf8]{inputenc}

\usepackage{upquote}

\begin{document}

\title{Evidence of weak circumstellar medium interaction in the Type II SN~2023axu}

\correspondingauthor{M. Shrestha}
\email{mshrestha1@arizona.edu}

\input{affiliation}
\input{author}








\begin{abstract}
We present high-cadence photometric and spectroscopic observations of SN~2023axu, a classical Type II supernova with an absolute $V$-band peak magnitude of $-16.5 \pm 0.1$ mag. SN~2023axu was discovered by the Distance Less Than 40 Mpc (DLT40) survey within 1 day of the last non-detection in the nearby galaxy NGC 2283 at 13.7 Mpc. We modeled the early light curve using a recently updated shock cooling model that includes the effects of line blanketing and found the explosion epoch to be MJD 59971.48 $\pm$ 0.03 and the probable progenitor to be a red supergiant with a radius of 417 $\pm$ 28 $R_\odot$. The shock cooling model cannot match the rise of observed data in the $r$ and $i$ bands and underpredicts the overall UV data which points to possible interaction with circumstellar material. This interpretation is further supported by spectral behavior. We see a ledge feature around 4600 \AA\ in the very early spectra (+1.1 and +1.5 days after the explosion) which can be a sign of circumstellar interaction. The signs of circumstellar material are further bolstered by the presence of absorption features blueward of H$\alpha$ and H$\beta$ at day $>$40 which is also generally attributed to circumstellar interaction. Our analysis shows the need for high-cadence early photometric and spectroscopic data to decipher the mass-loss history of the progenitor.

\end{abstract}



\section{Introduction} \label{sec:intro}
Massive stars with $M_{ZAMS} > 8 M_\odot$ end their lives with energetic explosions known as a core-collapse supernovae (CCSNe). Supernovae that exhibit hydrogen in their spectra are classified as type II\footnote{In this paper we use the term Type II to refer to both the Type IIP and IIL subtypes.} (SNe II) and are the most common type of CCSNe \citep{Li_2011,Smith_2011}. The progenitors of these explosions are red supergiant (RSG) stars, which has been confirmed by direct observations (see, e.g., \citealp{Smartt_2015,vandyk_2017}). 
However, there are still many open questions about SNe II progenitors including the mass-loss rate of the progenitor RSG during the last years prior to explosion \citep[e.g.][]{Ekstrom_2012, Beasor_2020, Massey_2023}. 

The mass-loss rate of RSGs in the months to years prior to explosion is difficult to observe directly. However, the circumstellar material (CSM) created by this mass loss can have an impact on both the light curve and spectra of SNe II at very early times \citep{Smith_2014}. In the absence of CSM interaction, the early light-curve evolution can be modeled considering shock breakout physics and subsequent cooling \citep[e.g.][]{Rabinak_2011, Sapir_2017, Morang_2023}. However, when such models have been implemented they do not match the early light curve evolution, often with discrepancies in the light curve rise and/or the bluest filters \citep[e.g.][]{Hosseinzadeh_2018,Hosseinzadeh_2023_23ixf,Hosseinzadeh_2022_21yja,Tartaglia_2018,Andrews_2019,Dong_2021,Pearson_2023}. \citet{Morozova_2017, Morozova_2018} found that radiation-hydrodynamical models with dense CSM fit the majority of early light curves of SNe II better than those without material around the progenitor star, pointing to the prevalence of CSM interaction. In addition to light curves, early spectra within hours to days of explosion can include clues about the CSM interaction in the form of emission lines created by the ionization of surrounding CSM by photons from shock breakout known as `flash' \citep[e.g.][]{GalYam_2014,Yaron_2017,Tartaglia_2021,Bruch_2021,Terreran_2022,Bostroem_2023_23ixf,Bruch_2023, Jacobson-Galan_2023} or `ledge' features \citep[e.g.][]{Hosseinzadeh_2022_21yja,Pearson_2023,Bostroem_2023}. 

All these studies show the value of early observations of SNe II. The latest generation of transient surveys have been successful in gathering impressive photometric data, however, rapid spectroscopic follow-up has often been limited both in terms of quantity and signal-to-noise. To address this lag between discovery and science-grade spectroscopic observations, we have developed a Python wrapper ({\sc PyMMT}) that interfaces with an Application Programming Interface (API) for rapid, same-night follow-up using the 6.5-meter MMT telescope. The details of {\sc PyMMT} are presented in Appendix \ref{sec:appendix}. 

\begin{figure}
\includegraphics[width=\columnwidth]{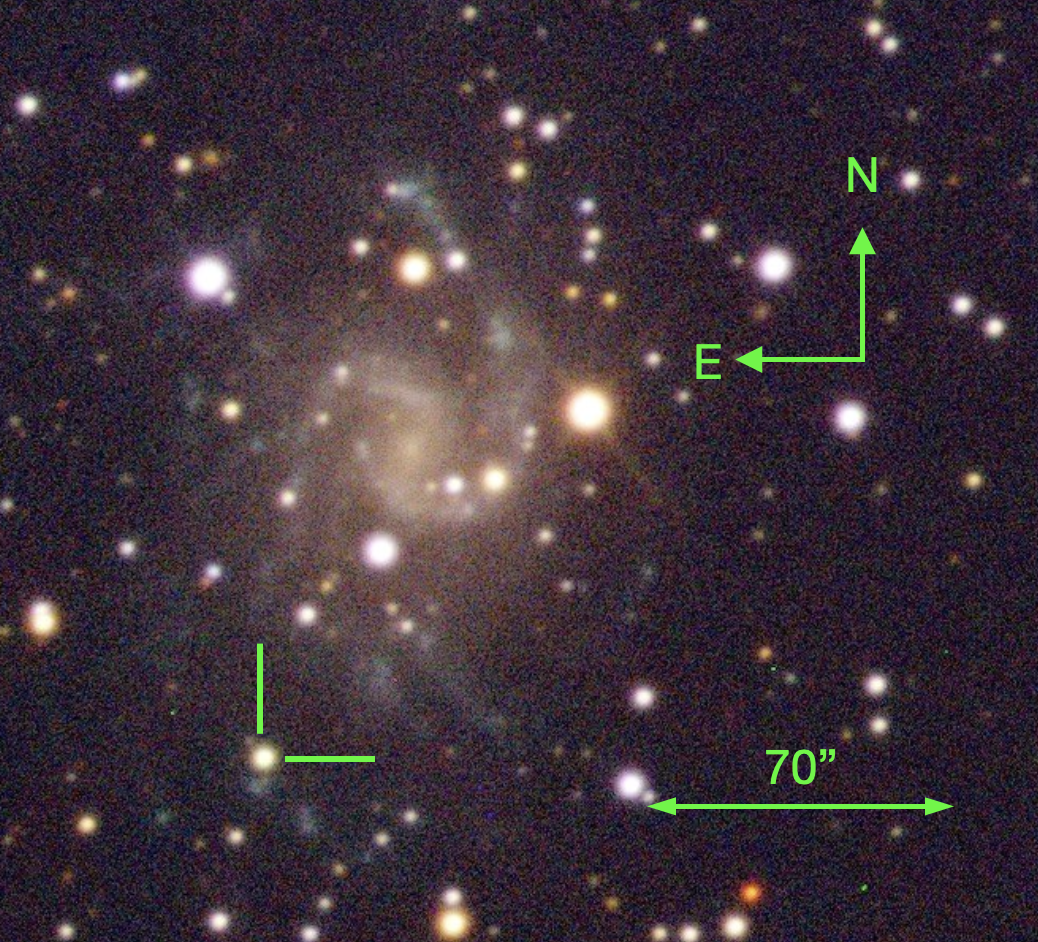}
\caption{Composite $g$, $r$, and $i$ image of SN~2023axu indicated by green tick marks in NGC 2283 obtained by Las Cumbres Observatory on 60058.37 MJD (+87 days after explosion).
\label{fig:galaxy_image}}
\end{figure}

In this paper, we discuss the observations and analysis of the Type II SN~2023axu. The SN~2023axu was first discovered by Distance Less Than 40 Mpc (DLT40) \citep{Tartaglia_2018} on 2023 January 28 (59972.11 MJD) with a discovery magnitude of 15.64 $\pm$ 0.01 in the clear filter \citep{Sand_2023} with the last non-detection on 59971.08 MJD from DLT40 \citep{Sand_2023} with a limiting magnitude of 20.04 mag. There is a later nondetection on 59971.517 MJD (priv. communication from K. Itagaki) with a limiting unfiltered magnitude of 19 mag. Asteroid Terrestrial-impact Last Alert System (ATLAS) first detected SN~2023axu on 59971.90 MJD a few hours prior to DLT40 discovery. DLT40 group reported the host galaxy to be NGC 2283 as shown in Fig.~\ref{fig:galaxy_image} along with SN~2023axu. The J2000 coordinates of the supernova are RA 06:45:55.32 and DEC -18:13:53.48. The SN was classified as a Type II SN by  \citet{Bostroem_2023tns} based on weak and broad hydrogen lines. The Type II classification was confirmed by the LiONS collaboration \citep{Li_2023} the following day. We made use of {\sc PyMMT} to trigger spectroscopic follow-up which resulted in an early spectrum within a day of discovery and +1.1 days after the explosion epoch. The properties of SN~2023axu are presented in Table.~\ref{tab:results}

The paper is organized as follows: in Section~\ref{sec:methods}, we present the observations and data reduction process along with the general properties of the supernova. In Section~\ref{sec:results}, we present our extinction calculations and the analysis we performed on the photometric data to calculate the nickel mass. We also compare our observed light curve with an updated shock-cooling model. Then we present the spectroscopic analysis showing the presence of a ledge feature and absorption lines on the blue side of H$\alpha$ and H$\beta$. We discuss and provide implications of our results and present conclusions in Section~\ref{sec:conclusions}.

\begin{table}
 \centering
 \caption{Properties of SN~2023axu} \label{tab:results}
 \begin{tabular}{ l c l}
    \hline
    Parameter & & Value \\
    \hline
    R.A. & & 06:45:55.32 (J2000) \\
    Dec. & & $-$18:13:53.52 (J2000)\\
    Last Non-Detection & &  59971.52 MJD\\ 
    First Detection & & 59972.12 MJD\\
    Explosion Epoch\footnote{from shock cooling fit} & & $59971.48 \pm 0.03$ MJD \\
    Redshift $z$ & & 0.002805 \\
    Distance & & $13.68 \pm 2.05$ Mpc\\
    Distance modulus ($\mu$) & & $30.68 \pm 0.32$ mag\\
    E$(B-V)_\mathrm{tot}$ & & $0.398 \pm 0.002$ mag\\
    Peak Magnitude ($V_{\mathrm{max}}$) & & $-16.53 \pm 0.15$ mag\\
    Time of $V_{\mathrm{max}}$ & & $59980.52 \pm 0.34$ MJD\\
    Nickel mass & & $0.029 \pm 0.010 M_\odot$ \\
    $s_{50}$ & & $0.37 \pm 0.01 $ mag/50 days \\
    Rise time ($V$) & & 8.9 days \\
    $t_{PT}$ & & $101.2 \pm 0.3$ days \\
    \hline
 \end{tabular}
\end{table}

\section{Observations and data reduction} \label{sec:methods}

\subsection{Host galaxy}
The host galaxy of SN 2023axu is NGC 2283 at a heliocentric redshift of $z = 0.002805 \pm 0.000005$ \citep{Koribalski_2004}. We assume the distance to the host galaxy to be $13.68\pm 2.05$ Mpc and a distance modulus of $30.68$ mag from the PHANGS survey \citep{Anand_2021}. For this distance calculation, \citet{Anand_2021} implements the numerical action methods (NAM) model described in \citet{Shaya_2017} and \citet{Kourkchi_2020}. The field of SN~2023axu was observed by ATLAS starting $\sim$ five years pre-explosion. We stacked the single-epoch flux measurements in 10-day bins following \cite{Young2022} to reach a deeper limit. There are no precursor outbursts observed with detection limits in $o$ and $c$ down to $\gtrsim -10.5$ mag. Most of the limits for SN~2023axu are fainter than the precursor of SN~2020tlf ($\gtrsim -11.5$ mag) \citep{Jacobson-Galan2022}.


\subsection{Photometric follow-up}
\begin{figure*}
\includegraphics[width=\textwidth]{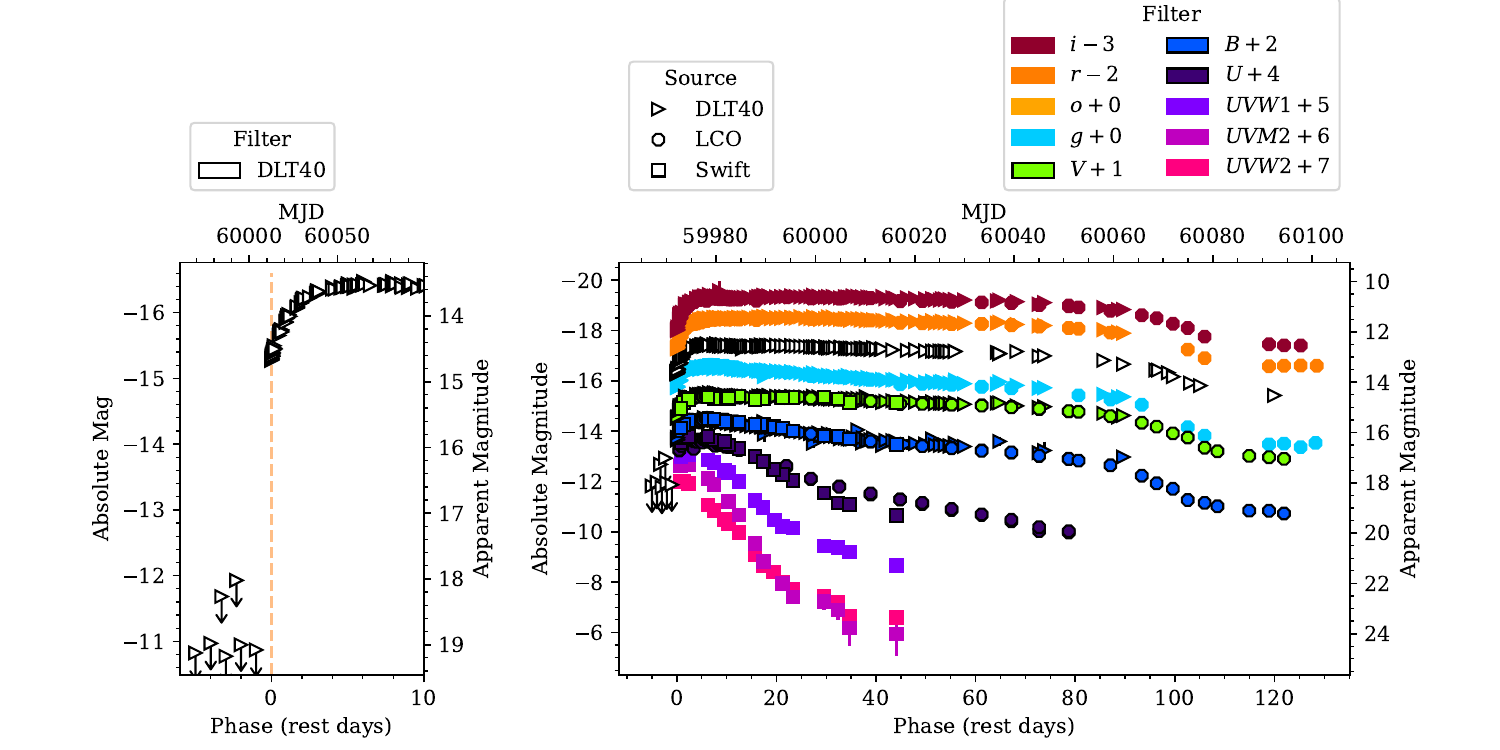}
\caption{Left: DLT40 clear filter zoom-in of the light curve for the first 10 days after the explosion.  Right: Multi-wavelength photometry in absolute and extinction-corrected apparent magnitudes of SN~2023axu spanning the UV to the optical observed using Swift, LCO, and DLT40. The light curve is well sampled throughout the first 150 days including the rise, plateau, fall from plateau and nickel tail. (The data used to create this figure are available in the published article.)
\label{fig:lc}}
\end{figure*}
After discovery, we continued imaging observations of SN 2023axu using DLT40 as well as the Las Cumbres Observatory network of 1-m telescopes \citep{Brown_2013} via the Global Supernova Project (GSP) with high cadence photometric observations. In addition, we also triggered the Ultraviolet/Optical Telescope on the Neil Gehrels Swift Observatory \citep{Gehrels_2004} and started observations on 59972.5 MJD a day after the discovery. 

Multi-band (BVgri) and Open filter data were taken utilizing SkyNet's network of 0.4m PROMPT telescopes \citep{Reichart_2005} via the DLT40 project. The BVgri filter data were pre-processed using a Python-based pipeline and aperture photometry was performed. These data are calibrated using the APASS catalog. Open filter data were template subtracted using HOTPANTS \citep{Becker_2015} and magnitudes were calibrated to r-band. 
The data from Las Cumbres Observatory were reduced using {\sc lcogtsnpipe} \citep{Valenti_2016}, a photometric reduction pipeline based on PyRAF. The APASS catalog was used to calibrate the BVgri filters and the Landolt catalog was used for U calibration. Finally, the Swift UVOT data were reduced following the prescription in \citet{Brown_2009} and the updated zero points from \citet{Breeveld_2011} were used for the calibration. The light curve from all the instruments is presented in Fig.~\ref{fig:lc}.

\subsection{Spectroscopic follow-up}
\begin{figure*}
\includegraphics[width=\textwidth]{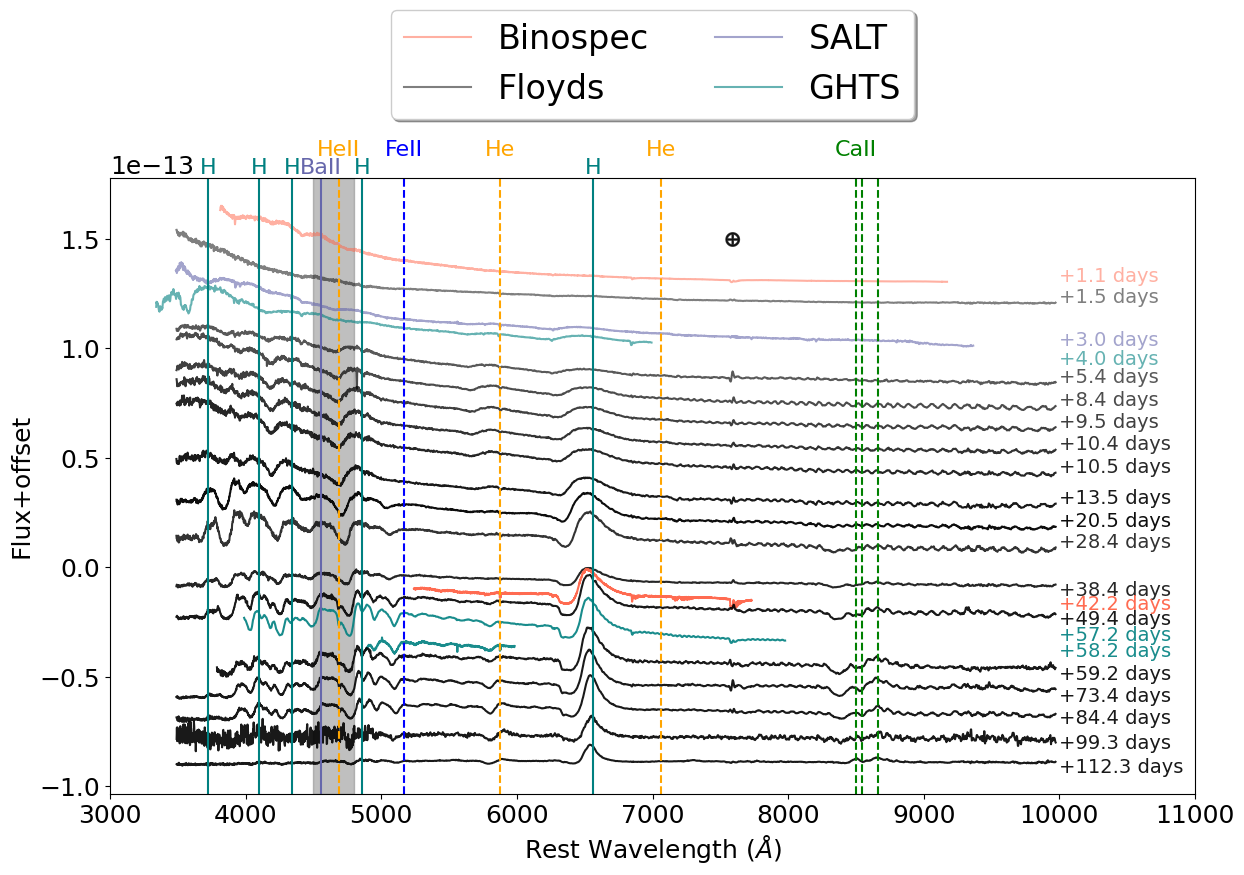}
\caption{Optical spectral evolution of SN~2023axu starting at $+$1.1 days after the explosion and ending at $+$112.3 days. Several spectral features are indicated by vertical lines and the `ledge' feature wavelength range is shown in the gray shaded region. All the spectra are extinction corrected. All spectra will be made available on WISeRep (\url{https://www.wiserep.org}).
\label{fig:spec_all}}
\end{figure*}

We observed SN 2023axu spectroscopically using various facilities: FLOYDS on Faulkes Telescope North \citep[FTN; ][]{Brown_2013} as a part of the GSP, Binospec \citep{Fabricant_2019} on the MMT on Mt. Hopkins AZ, the Boller and Chivens Spectrograph (B\&C) on the Bok 2.3m telescope located at Kitt Peak National Observatory, the Robert Stobie Spectrograph (RSS) on the Southern African Large Telescope \citep[SALT;][]{SALT}, and the Goodman High-Throughput Spectrograph \citep[GHTS;][]{soar} on the Southern Astrophysical Research Telescope (SOAR; 4.1~m telescope at Cerro Pachon, Chile). We used our newly developed PyMMT wrapper (described in Appendix~\ref{sec:appendix}) to trigger rapid spectroscopy with MMT Binospec, obtaining our first spectrum +1.1 days after explosion and the same day as the discovery. A log of spectroscopic observations is presented in Table ~\ref{tab:specInst} and the spectral evolution is presented in Fig.~\ref{fig:spec_all}.

The FLOYDS data were reduced using a purpose built pipeline in IRAF \citep{Valenti_2014}. For the Binospec data, the initial data processing of flat-fielding, sky subtraction, and wavelength and flux calibration was done using the Binospec IDL pipeline \citep{BinoIDL}\footnote{\url{https://bitbucket.org/chil_sai/binospec/wiki/Home}}. We then extracted the 1D spectrum using IRAF \citep{iraf1,iraf2}. The B$\&$C spectra were reduced using  standard IRAF reduction techniques. The RSS spectra from SALT were reduced using a custom longslit pipeline based on the PySALT package \citep{PySALT}. Finally, Goodman spectra were reduced using a custom Python package developed by SOAR Observatory\footnote{https://soardocs.readthedocs.io/projects/goodman-pipeline/en/latest/}.

\begin{table*}
 \centering
 \caption{Log of Spectroscopic Observations}
 \begin{tabular}{ c c c  c c c c}
    \hline
    Date (UTC) & MJD &  Telescope & Instrument & Range ({\AA}) & Exp (s) & Slit ('')\\
    \hline
    2023-01-28 & 59972.171 &  MMT & Binospec & 3900--9240 & 3600 & 1\\
    2023-01-28 & 59972.525 &   FTS & FLOYDS & 3500--10000 & 1800 & 2\\
    2023-01-30 & 59974.085 &   SALT & RSS & 3495--9390 & 1893 & 1.5\\ 
    2023-01-31 & 59975.085 &   SOAR & GHTS RED & 3350--7000 & 285 & 1\\ 
    2023-02-01 & 59976.466 &  FTS & FLOYDS & 3500--10000 & 1800 & 2\\
    2023-02-04 & 59979.431 &   FTS & FLOYDS & 3500--10000 & 1799 & 2\\
    2023-02-05 & 59980.539 &  FTS & FLOYDS & 3500--10000 & 1800 & 2\\
    2023-02-06 & 59981.489 &   FTS & FLOYDS & 3500--10000 & 1200 & 2\\
    2023-02-06 & 59981.534 &   FTS & FLOYDS & 3500--10000 & 1200 & 2\\
    2023-02-09 & 59984.603 &   FTS & FLOYDS & 3500--10000 & 1200 & 2\\
    2023-02-16 & 59991.538 &   FTS & FLOYDS & 3500--10000 & 1200 & 2\\
    2023-02-24 & 59999.488 &   FTS & FLOYDS & 3500--10000 & 900 & 2\\
    2023-03-06 & 60009.445 &   FTS & FLOYDS & 3500--10000 & 900 & 2\\
    2023-03-10 & 60013.445 &   MMT &  Binospec & 5255--7753 & 1800 & 1\\
    2023-03-17 & 60020.410 &   FTS & FLOYDS & 3500--10000 & 900 & 2\\
    2023-03-25 & 60028.249 &   Bok & B\&C & 4100--8000 & 900 & 1.5\\
    2023-03-26 & 60029.249 &   Bok & B\&C & 4100--8000 & 900 & 1.5\\
    2023-03-27 & 60030.249 &   FTS & FLOYDS & 3500--10000 & 900 & 2\\
    2023-04-10 & 60044.430 &   FTS & FLOYDS & 3500--10000 & 1500 & 2\\
    2023-04-21 & 60055.390 &   FTS & FLOYDS & 3500--10000 & 1500 & 2\\
    2023-05-06 & 60070.353 &   FTS & FLOYDS & 3500--10000 & 1500 & 2\\
    2023-05-19 & 60083.356 &   FTS & FLOYDS & 3500--10000 & 1800 & 2\\
    \hline
 \end{tabular}
 
 \label{tab:specInst}
\end{table*}


\section{Analysis} \label{sec:results}

\subsection{Extinction}
 The equivalent width of the Na ID absorption line is known to correlate with the interstellar dust extinction \citep{Richmond_1994, Munari_1997,Poznanski_2012}. We used an MMT Binospec spectrum with a resolution of 1340 ($\lambda/\delta \lambda$) to fit the equivalent width to the Na I D1 and Na I D2 absorption features which are clearly separated. We simultaneously fit a Gaussian function to the absorption of both the Milky Way and the host galaxy along with the continuum using the astropy modeling package \citep{astropy:2013,astropy:2018,astropy:2022}. The best fit equivalent width was used in Equation 9 from \citet{Poznanski_2012} and a renormalization factor of 0.86 from \citet{Schlafly_2011} was applied. This gives  $E(B-V) = 0.383 \pm 0.015$ mag for the Milky Way extinction (Note: this value of extinction is on the upper end of \citet{Poznanski_2012} sample distribution). This value is in agreement with  Milky Way extinction of $E(B-V) = 0.3319 \pm 0.0107$ from \citet{Schlafly_2011} from the IPAC Dust Service \footnote{https://irsa.ipac.caltech.edu/applications/DUST/} at the SN~2023axu co-ordinates. We find a low level of extinction for the host galaxy with $E(B-V) = 0.015 \pm 0.001$ mag. In this paper, we use the Milky Way extinction value calculated using the Na ID absorption line, thus the total color excess is $E(B-V) = 0.398 \pm 0.015 $ mag.

\subsection{Photometry}
\begin{figure}
    \centering
    \includegraphics[width=\columnwidth]{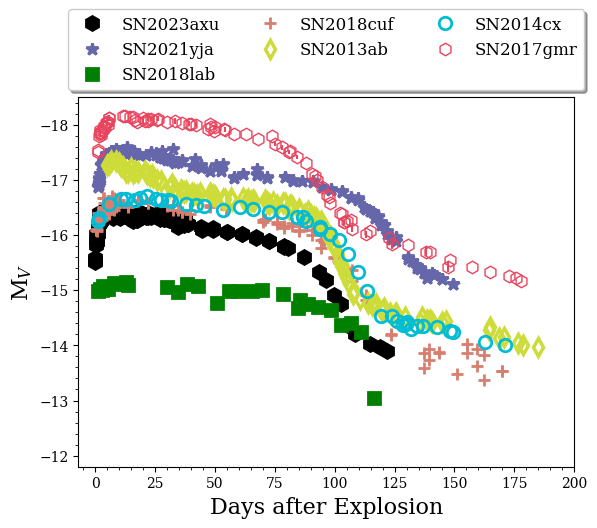}
    \caption{Absolute $V$-band magnitude of SN~2023axu (black) compared to other Type II SNe: SN~2021yja \citep{Hosseinzadeh_2022_21yja}, SN~2018lab \citep{Pearson_2023}, SN~2018cuf \citep{Dong_2021}, SN~2013ab \citep{Bose_2015}, SN~2014cx \citep{Huang_2016}, SN~2017gmr \citep{Andrews_2019}. The peak $V$-band magnitude of SN~2023axu is -16.53 mag which is closest to SN~2018cuf and SN~2014cx. }
    \label{fig:lccomp}
\end{figure}

The multi-wavelength light curve of SN~2023axu is presented in Figure~\ref{fig:lc}. The light curve evolves as a normal SN II which can be seen in Fig.~\ref{fig:lccomp}. We perform analysis on SN~2023axu photometric data as done in \citet{Valenti_2016} for a sample of SN II to characterize the light curve. The peak brightness is reached between 59979.59 MJD (+8.2 days) and 59980.92 MJD (+9.5 days), depending on the filter. It has a normal rise time of 8.9 days in the $V$ band for the peak magnitude of $-16.53$ for an SNe II as seen in \citet[Fig.16 (top)]{Valenti_2016}

\begin{figure}
    \centering
    \includegraphics[width=\columnwidth]{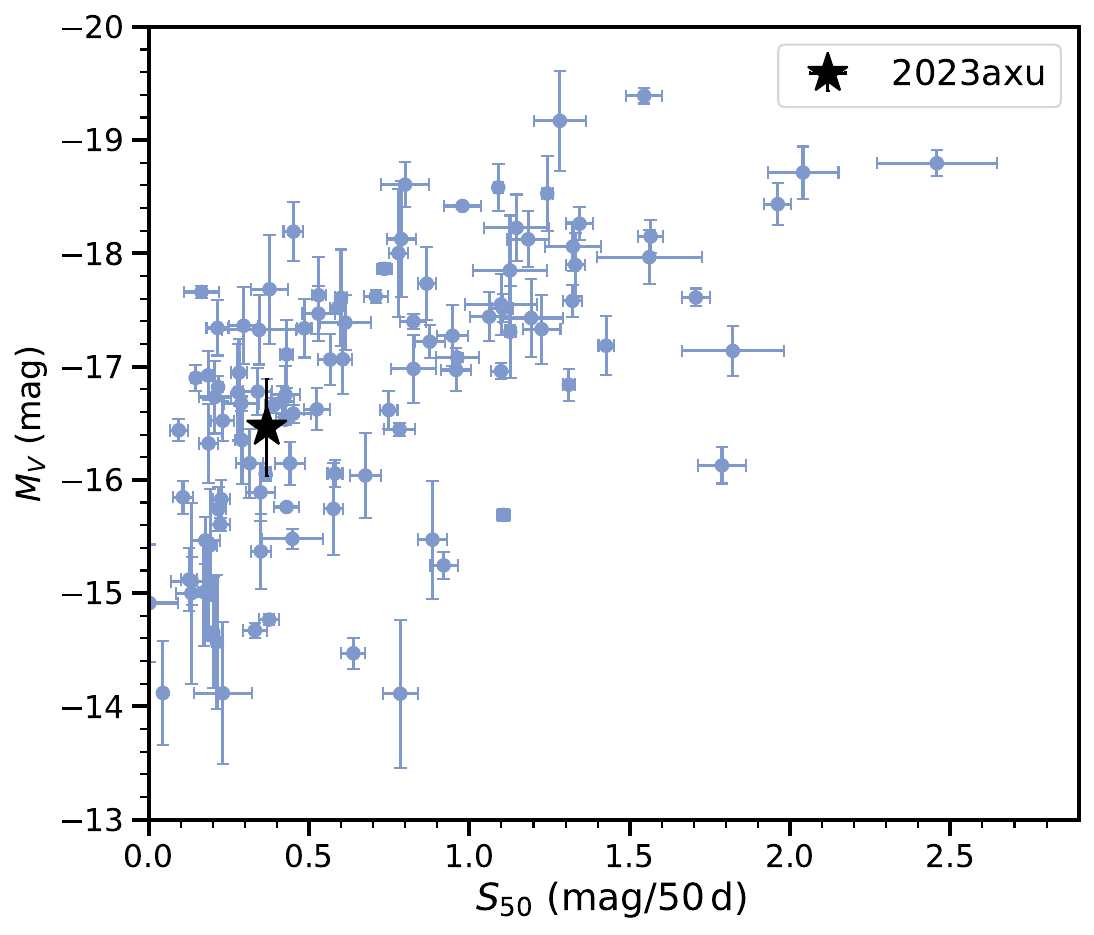}
    \caption{Comparison of V-band peak absolute magnitude with respect to slope (mag/(50 days)) with other SNe II from \citet{Valenti_2016}. The value of SN~2023axu falls in the normal part of the trend seen by \citet{Valenti_2016}.}
    \label{fig:peakslope}
\end{figure}

The $V$-band decline rate in the 50 days after the light curve reaches the plateau is denoted by $s_{50}$ and calculated following the prescription of \citet{Valenti_2016}. For SN~2023axu we find $ s_{50} =0.37 \pm 0.01 $ mag/50 days. In Figure~\ref{fig:peakslope}, we plot the peak in $V$-band absolute magnitude against  $s_{50}$ for type SNe II including SN~2023axu which falls in the normal range for other SNe of the same class. As expected for normal SNe II, the rise is followed by a plateau phase ($t_{PT}$, described in \citet{Valenti_2016}) which lasts for $101.2 \pm 0.3$ days for SN~2023axu. This value is also in the normal range for a SNe II of similar magnitude. In the next section, we use data after the fall from the plateau to calculate the nickel mass.

\subsection{Nickel mass}

\begin{figure}
\includegraphics[width=\columnwidth]{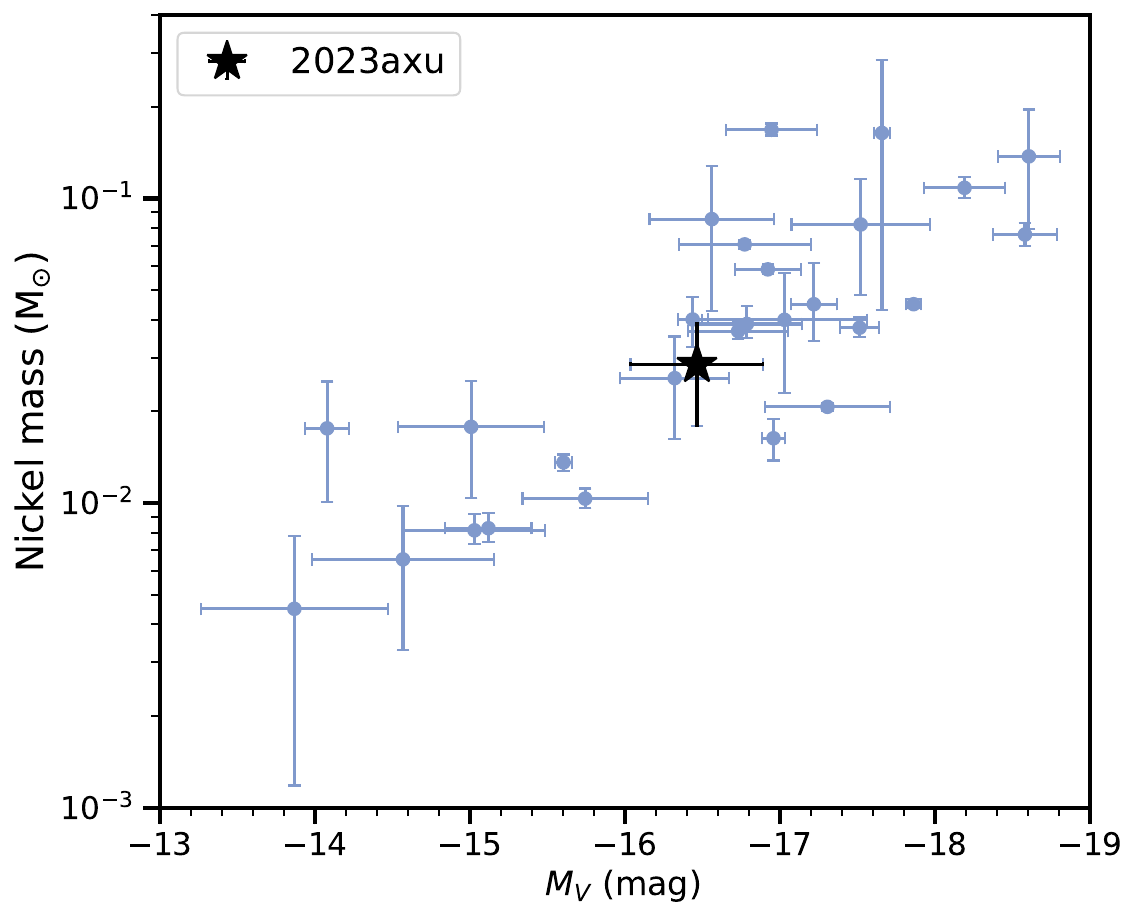}
\caption{Nickel mass with respect to $V$-band absolute mag at 50 days for SN~2023axu (star) along with other SNe II from \citet{Valenti_2016}. The value of the SN follows the expected trend.
\label{fig:nickel_comp}}
\end{figure}

Towards the end of the dataset presented in this paper, SN~2023axu falls from plateau, 
as shown in Figure~\ref{fig:lc}, and settles on the radioactive-decay tail. This phase of the SNe II is driven by the radioactive decay of nickel to iron ($\rm ^{56}Ni\rightarrow{}^{56}Co\rightarrow{}^{56}Fe$). To calculate the nickel mass for SN~2023axu, we use data in $B,V,g,r,i$ filters. We calculate pseudo-bolometric luminosity following \citet{Valenti_2008}. We then compare this pseudo-bolometric luminosity with the pseudo-bolometric light curve of SN~1987A ($B,V,g,r,i$) (we note that for $g,r,i$ Sloan filters, we perform synthetic photometry on SN~1987A spectra) following the method by \citet{Spiro_2014}; $M_{Ni} = 0.075 M_\odot \times \frac{L_{23axu}}{L_{87A}}$, where $L_{23axu}$ and $L_{87A}$ are the pseudo-bolometric luminosity of SN~2023axu and SN~1987A respectively. For this calculation to be valid we assume the ejecta completely traps the $\gamma$-rays produced by the radioactive decay.  The pseudo-bolometric curve for this period declines similarly to that of fully-trapped $^{56}Co$ decay of 0.98 mag 100 days$^{-1}$.
To calculate $L_{23axu}$ and $L_{87A}$, we make use of data from 105 days after the explosion epoch. Finally, we substitute the calculated $L_{23axu}$ and $L_{87A}$ in  $M_{Ni} = 0.075 M_\odot \times \frac{L_{23axu}}{L_{87A}}$ and find $M_{Ni} = 0.0285^{+0.01}_{-0.01} M_\odot$. Both \citet{Valenti_2016} and \citet{ Anderson_2014} found a relation between nickel mass and absolute magnitude in the V band at 50 days for their SNe II samples. We overplot the value of SN~2023axu in the sample by \citet{Valenti_2016} and find it to follow the trend as shown in Figure~\ref{fig:nickel_comp}.

\subsection{Shock cooling model}\label{sec:lcmodel}
\begin{deluxetable*}{lCcCCCCc}
\label{tab:shockcooling}
\tablecaption{Shock-cooling Parameters\label{tab:params}}
\tablehead{&& \multicolumn{3}{c}{Prior} & \multicolumn{2}{c}{Best-fit Values\tablenotemark{a}} & \\[-10pt]
\colhead{Parameter} & \colhead{Variable} & \multicolumn{3}{c}{------------------------------------------} & \multicolumn{2}{c}{------------------------------------} & \colhead{Units} \\[-10pt]
&& \colhead{Shape} & \colhead{Min.} & \colhead{Max.} & \colhead{MSW23} & \colhead{SW17} & }
\startdata
Shock velocity                        & v_\mathrm{s*}  & Uniform & 0.2      & 1.5    & 1.14^{+0.07}_{-0.06}        & 0.87^{+0.04}_{-0.03}        & $10^{8.5}$ cm s$^{-1}$ \\
Envelope mass\tablenotemark{b}        & M_\mathrm{env} & Uniform & 1      & 3    & 1.2^{+0.2}_{-0.1}              & 1.2\pm 0.1        & $M_\sun$ \\
Ejecta mass $\times$ numerical factor & f_\rho M       & Uniform & 0    & 1.2   & 0.6^{+0.2}_{-0.1}        & 0.7 \pm 0.2                & $M_\sun$ \\
Progenitor radius                     & R              & Uniform & 0      & 1438 & 417 \pm 28               & 560\pm{+43}        & $R_\sun$ \\
Explosion time                        & t_0            & Uniform & 59970 & 59972.7  & 59971.48 \pm 0.03  & 59971.26 ^{+0.01}_{-0.02} & $\mathrm{MJD}$ \\
Intrinsic scatter                     & \sigma         & Log-uniform & 0  & 10^2  & 5.8 \pm 0.2             & 5.7 \pm 0.2            & \nodata \\
\enddata
\tablenotetext{a}{The ``Best-fit Values'' columns are determined from the 16th, 50th, and 84th percentiles of the posterior distribution, i.e., $\mathrm{median} \pm 1\sigma$. 
MSW23 and SW17 stand for the two models from \cite{Morang_2023} and \cite{Sapir_2017}, respectively. The former is preferred.}
\tablenotetext{b}{See Section~\ref{sec:lcmodel} for the definition of ``envelope'' in the shock-cooling paradigm.}
\end{deluxetable*}
\vspace{-24pt}

The early photometric evolution of SNe II is thought to be driven by shock cooling emission, at least in the absence of significant CSM. This form of emission carries the signatures of the progenitors. Hence, modelling this emission using analytic recipes can play an important role in constraining the properties of progenitor stars. With this in mind, we utilized the shock-cooling model of \citet{Morang_2023} (hereafter; MSW23) to model the early light curve of SN~2023axu. We fit this model using the Light Curve Fitting package \citep{hosseinzadeh_light_2023a}. Recently, this model has been successfully implemented for the case of SN~2023ixf \citep{Hosseinzadeh_2023_23ixf} where a good match to the early light curve was found. In this paradigm, the star is assumed to be a polytrope with a density profile $\rho_0 = \frac{3 f_\rho M}{4\pi R^2} \delta^n$, where $f_\rho$ is a numerical factor of order unity, $M$ is the ejecta mass (with the remaining remnant neglected), $R$ is the stellar radius, $\delta \equiv \frac{R-r}{R}$ is the fractional depth from the stellar surface, and $n = \frac{3}{2}$ is the polytropic index for convective envelopes. The shock velocity profile is described by $v_\mathrm{sh} = v_\mathrm{s*} \delta^{-\beta n}$, where $v_\mathrm{s*}$ is a free parameter and $\beta = 0.191$ is a constant. In the shock-cooling model, we treat $f_\rho M$ as a single parameter because they always appear together and are highly degenerate. The unknown core-collapse explosion time is parameterized by $t_0$. $M_\mathrm{env}$ is the mass in the stellar envelope, defined as the region where $\delta \ll 1$. Finally, we also include an intrinsic scatter term $\sigma$ which accounts for the scatter around the model as well as probable underestimates of photometric uncertainties. We multiply the observed error bars by a factor of $\sqrt{1+\sigma^2}$. The \citet{Morang_2023} model is built on previous shock-cooling models \citep{Sapir_2011, Sapir_2013, Katz_2012, Rabinak_2011,Sapir_2017}. The model by \citet{Sapir_2017} (hereafter; SW17) has identical fit parameters as \citet{Morang_2023} with a few key differences. First, they do not account for the very early phase where the thickness of the emitting shell is smaller than the stellar radius, second, they assume a blackbody SED at all times whereas \citet{Morang_2023} account for some line blanketing in UV. We fit our observed light-curve data with both SW17 and MSW23 models and compare the results for completeness.

\begin{figure*}
\includegraphics[width=3.5in]{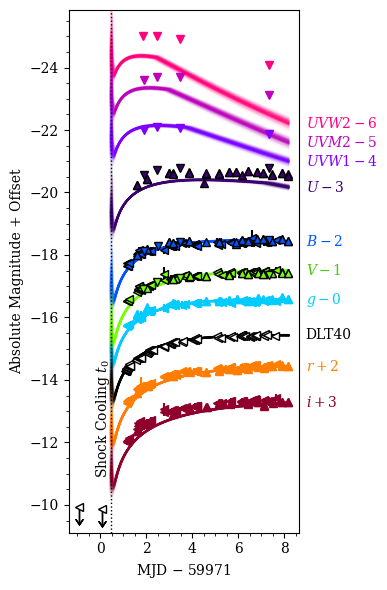}
\includegraphics[width=3.5in]{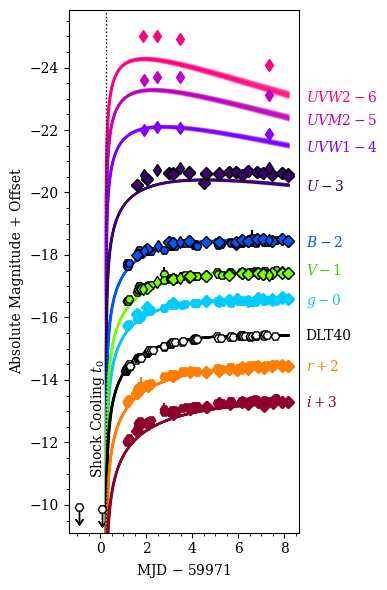}
\caption{Shock cooling modeling of SN~2023axu using the prescription of \citet{Morang_2023} (left) and \citet{Sapir_2017} (right). The best-fit explosion epoch for each model is shown by a dotted line. 
\label{fig:shockcooling}}
\end{figure*}

The result of the MSW23 and SW17 shock-cooling model for the early light curve of SN~2023axu is presented in Fig.~\ref{fig:shockcooling} and the best-fit parameters are presented in Table~\ref{tab:shockcooling}. We find that both models converge and give an overall good fit with some significant discrepancies for the observed data.  We find the best-fit parameters are reasonably comparable between the two prescriptions. However, the progenitor radius is different which has been seen previously by \citet{Hosseinzadeh_2023_23ixf} for SN~2023ixf. We find the error in radius estimate to be lower for MSW23 compared to SW17. 

The best-fit model for MSW23 constrains the radius of the progenitor to be $R = 417 \pm 28\ R_\sun$, which falls in a reasonable range for RSGs (100-1500 $R_\sun$; \citet{Levesque_2017}). We also constrain the explosion time to be $t_0 =  59971.48 \pm 0.03$ MJD which is $\sim 0.7$ days before the discovery detection by DLT40, and after the last DLT40 non-detection 59971.084 MJD. We note the last non-detection from Itagaki is inconsistent with the explosion time from the shock cooling model and they would have made a detection at $59971.48\pm 0.03$ MJD.  All the best parameters using \citet{Morang_2023} along with comparisons to \citet{Sapir_2017} model are presented in Table~\ref{tab:params}.

We find some discrepancies in $i,r$ and different $UV$ filters. To our knowledge, this is the first time the shock-cooling model of \citet{Morang_2023} has been used for a full set of data including $UV$ filters. \citet{Hosseinzadeh_2023_23ixf} were not able to perform this for a full set of $UV$ data due to the brightness and proximity of SN~2023ixf which saturated many of the $UV$ detectors. For SN~2023axu, the $i,r$ best-fit does not rise as quickly as the data. \citet{Morozova_2017, Morozova_2018} have shown that models without CSM cannot predict this rise well. In their model, the presence of dense CSM provides a better fit to the first $\sim$ 20 days of the SNe II light curves. Thus, the excess in the $i,r$ bands compared to the shock-cooling model could be explained by the presence of CSM. We find that the best-fit data under-predict our observed data for $U, UVM2$, and $UVW2$ filters for all epochs. For $UVW1$, the model also underpredicts our observations for later epochs, the phases for which MSW23 modify the SED to account for UV line blanketing. Therefore this discrepancy in the $UV$ filters for SN~2023axu could be due to \citet{Morang_2023} over-correcting for the line blanketing in the $UV$.  

So far, we do not have any cases where the $UV$ data fits the predictions from the shock-cooling models -- recent studies include SN~2018cuf \citep{Dong_2021}, SN~2017gmr \citep{Andrews_2019}, and SN~2016bkv \citep{Hosseinzadeh_2018}, although these efforts have used the previous version of the shock-cooling prescription \citep{Sapir_2017}.  A systematic effort to model early SN II light curves with the latest shock cooling models is warranted.

\subsection{Spectral evolution}
We present the optical spectral evolution of SN~2023axu in Fig.~\ref{fig:spec_all} spanning from 1.1 days to 112.3 days after the explosion (using the estimated explosion epoch derived from light-curve fitting in Section~\ref{sec:lcmodel}). The spectral evolution of SN~2023axu is largely typical for a SNe II with P Cygni lines of hydrogen and helium that develop $\sim 10$ days after explosion. We can see the formation of Ba II, Fe II, and Ca II lines at later stages. However, we see two interesting features in our spectra; 1) at early times, before $\sim$3 days with respect to explosion, we see a `ledge-shaped' feature around $4500 - 4800$~\AA, and 2) there is a flat absorption component in the H$\alpha$ P Cygni profile, along with an additional shallow absorption feature. We discuss these aspects of the spectral evolution in detail below.

\subsubsection{Ledge feature}
\begin{figure*}
\includegraphics[width=\textwidth]{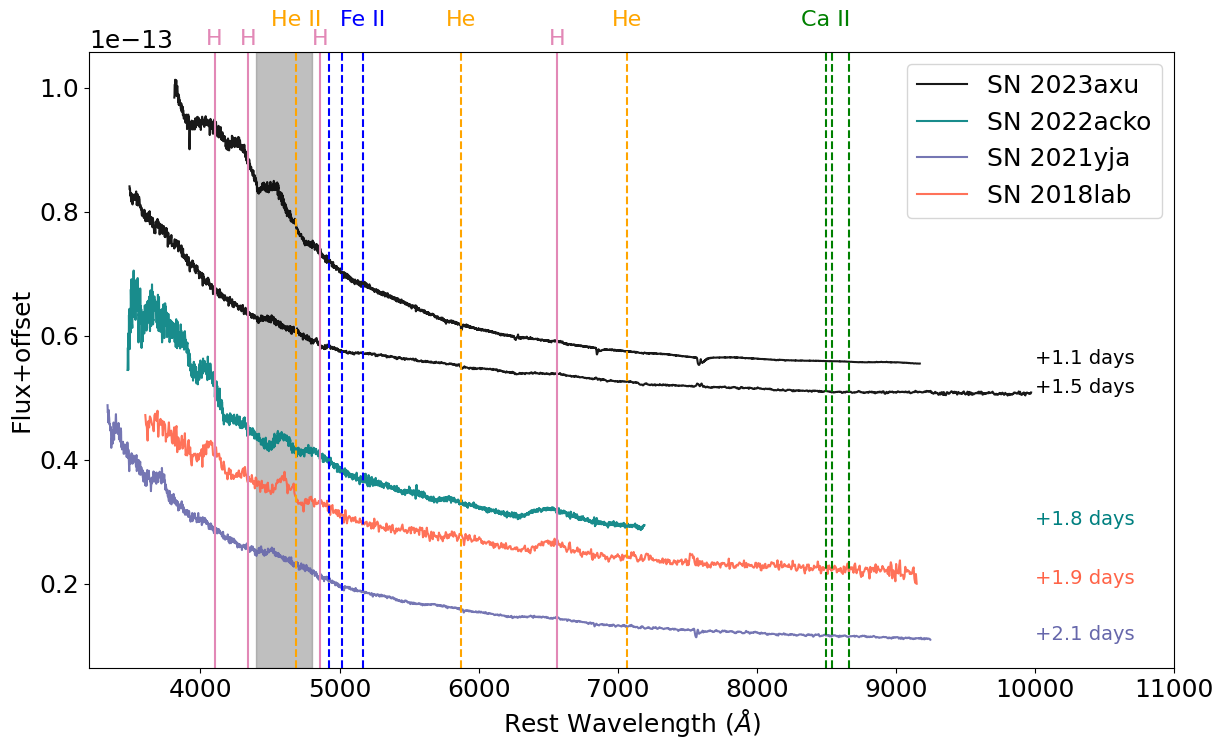}
\includegraphics[width=\textwidth]{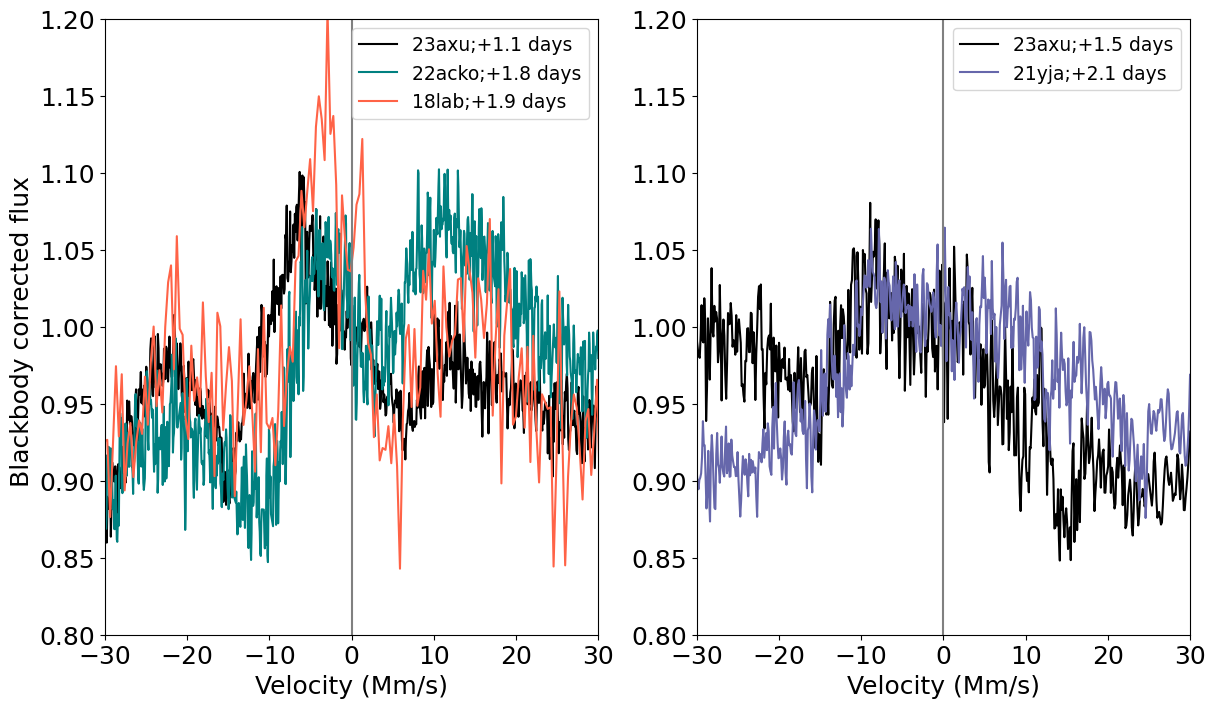}
\caption{(top) First two epochs of SN~2023axu compared to several other SNe II with observations of 'ledge' features, highlighted by the gray shaded region.
(bottom) Ledge feature present around $4600$ \AA~for SN~2023axu compared to SN~2022acko and SN~2018lab at +1.1 days (left) and SN~2021yja at +1.5 days (right). 
\label{fig:ledge_comp}}
\end{figure*}

The `ledge' feature seen around $4600$~\AA~and spanning roughly from $4400$~\AA~to $4800$~\AA~ (Fig.~\ref{fig:ledge_comp}) is present in the first two epochs of our spectroscopic observations. There is no clear signature of this feature starting +3 days after explosion. This ledge feature has been observed in a few other supernovae, for example: SN~2017gmr \citep{Andrews_2019}, SN~2018fif \citep{Soumagnac_2020}, SN~2018lab \citep{Pearson_2023}, SN~2021yja \citep{Hosseinzadeh_2022_21yja}, and SN~2022acko \citep{Bostroem_2023}. In the literature, this feature has been attributed to circumstellar interactions, however, the interpretation of this interaction is explained in two different ways; 
\begin{enumerate}
    \item \citet{Bullivant_2018, Andrews_2019} explained it as a broad, blueshifted He II 4686\AA~line that is produced in the outermost layer of SN ejecta beneath a CSM shell.
    \item Other papers have attributed this feature to bulk motion creating broad lines and blending of several ionized features from the CSM \citep{Soumagnac_2020, Bruch_2021}.
\end{enumerate}

The observation of the ledge feature and lack of prominent emission line flash features in SN~2023axu adds to the spectral diversity of SNe II, although we note the possibility of the presence of flash features in the spectra before our first epoch of observation. In Fig.~\ref{fig:ledge_comp} top panel, we plot the first two epochs of SN~2023axu along with three other SNe II showing the ledge feature.
We zoom in on the ledge feature in the bottom panel, where we have plotted a continuum normalized flux for this feature in comparison to a few other SNe for the first and second epoch of spectroscopic observations (+1.1 days and +1.5 days respectively). We find that for the first epoch, the ledge feature of SN~2023axu behaves similarly to that seen in SN~2018lab and SN~2022acko  (Fig.~\ref{fig:ledge_comp}, left) both of which are low luminosity SNe II. For the second epoch, the shape of the feature is very similar to that seen for SN2021yja, a $UV$-bright SN II (Fig.~\ref{fig:ledge_comp}, right). This implies that the diversity of this feature may be due to the rapid temporal evolution of which we only have sparse sampling rather than the intrinsic properties of the SN itself.


\begin{figure*}
\includegraphics[width=\textwidth]{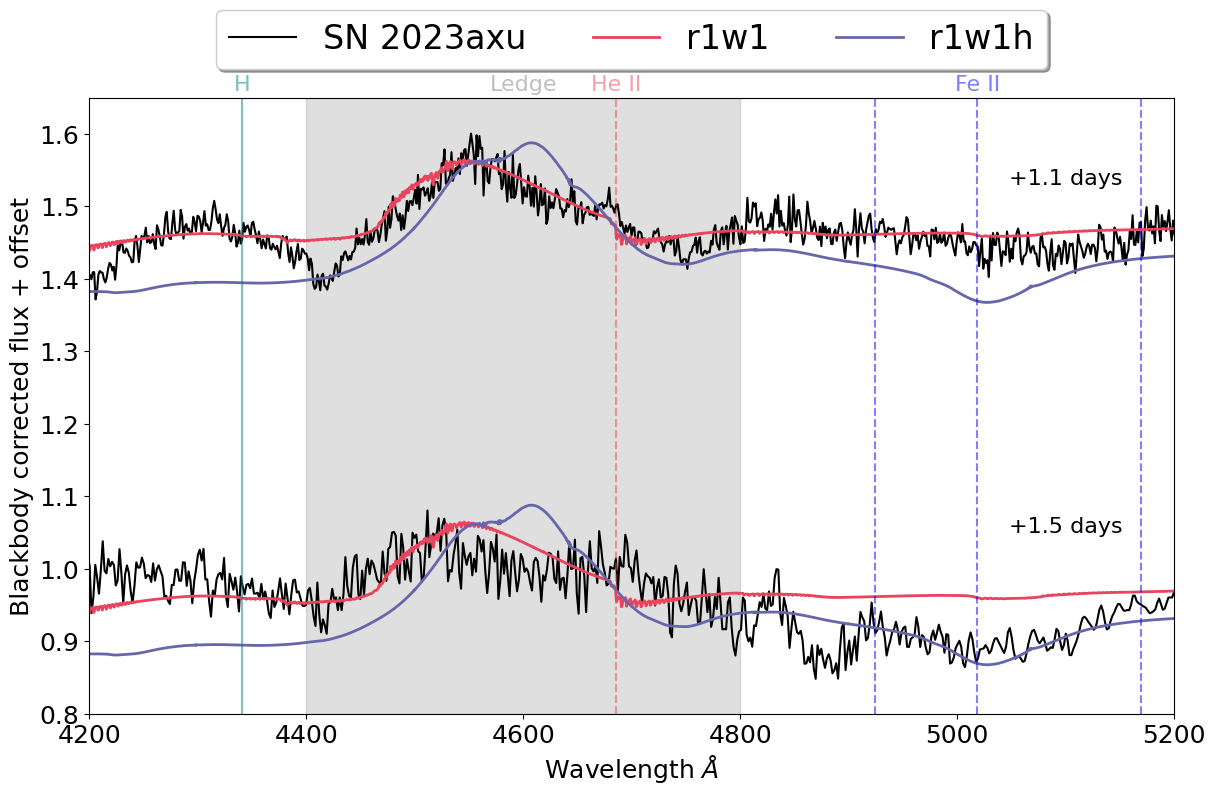}
\caption{ Comparison of the ledge feature of SN~2023axu at +1.1 days and +1.5 days with models at +1.0 days for \texttt{r1w1} and +1.6 days for \texttt{r1w1h} by \citet{Dessart_2017}.   Both models are for RSGs with CSM and model \texttt{r1w1h} is with extended atmospheres. The grey-shaded region indicates the approximate wavelength range for the ledge feature.
\label{fig:ledge_comp_model}}
\end{figure*}

We also compared this feature with radiation transfer simulations by \citet{Dessart_2017} where they model spectra and light curves for a RSG with a radius of $R_{\star} = 501 R_{\odot}$ which explodes in a low-density CSM of $\dot{M} = 10^{-6} M_{\odot}$ yr$^{-1}$ (the  \texttt{r1w1} model). The same model as \texttt{r1w1} with the addition of an extended atmosphere with a scale height of $H\rho = 0.3 R_{\star}$ is called \texttt{r1w1h}. The comparison of the first two epochs of spectra showing the ledge feature to the \texttt{r1w1} and \texttt{r1w1h} models is shown in Fig.~\ref{fig:ledge_comp_model}. 
There is no model that fits both of the epochs well. The peak of the \texttt{r1w1h} model is distinct from any feature seen in SN~2023axu, however, the model \texttt{r1w1} seems to be qualitatively better matched to the morphology of the SN~2023axu ledge feature seen at +1.1 days. This preference suggests that the CSM around SN~2023axu is of low density, which may be the reason why we do not see narrow emission line flash features.   

\subsubsection{\texorpdfstring{H$\alpha$}{Ha} and Cachito feature}
The P Cygni H$\alpha$ feature develops $\sim$ 10 days after explosion as seen in Fig.~\ref{fig:spec_all}. The H$\alpha$ absorption feature broadens and develops into a square trough after $\sim$28 days. The ratio of the equivalent width of the absorption to emission ($\rm a/e$) components at +28.4 days is 0.19. We also calculate the H$\alpha$ velocity (full-width half maximum of the emission) at the same epoch and found it to be 8770 $\rm km s^{-1}$. The relation between $\rm a/e$ with respect to H$\alpha$ velocity has been explored in \citet{Gutierrez_2014} and found the cases with smaller $\rm a/e$ have higher velocities. We compared the value from SN~2023axu with their sample and found it to fall in the normal range.

\begin{figure}
\includegraphics[width=\columnwidth]{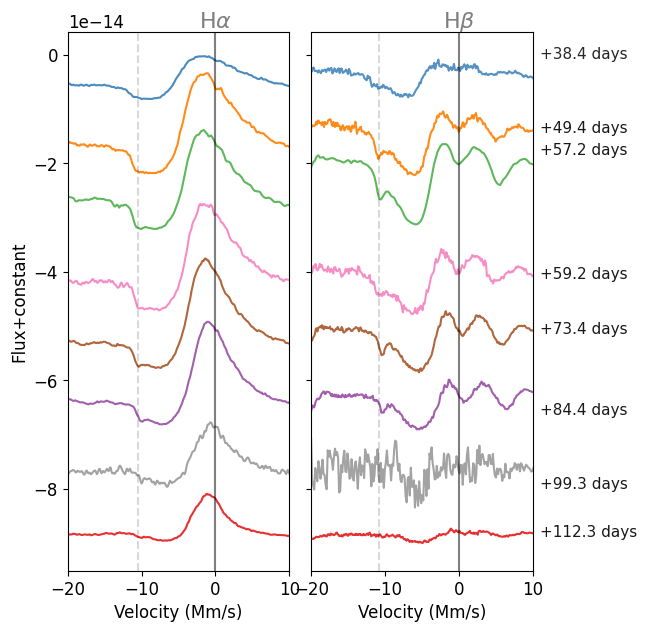}
\caption{Evolution of the Cachito feature for H$\alpha$ (left) and H$\beta$ (right) shown by the dashed vertical line. The velocity is calculated with respect to H$\alpha$ 6562.8 \AA~and H$\beta$ 4861 \AA~shown with gray lines. The shallow absorption feature is seen for both H$\alpha$ and H$\beta$ at a very similar velocity.
\label{fig:cachito}}
\end{figure}

There is an additional shallow absorption line of unclear origin within this broad feature at $\sim 10000$ $\rm km s^{-1}$, starting at +49 days. Similar absorption components have been seen in other supernovae such as  SN~2007X, SN~2004fc, SN~20023hl, SN~1992b, \citep{Gutierrez_2017}, and SN~2020jfo \citep{Teja_2022}, and is often dubbed the `Cachito' feature \citep{Gutierrez_2017}. 

From a large sample study, \citet{Gutierrez_2017} found that SNe with a Cachito line can be divided into two groups, depending on when the feature is present. In the first group, it is seen around 5-7 days between 6100 and 6300 \AA~and disappears at $\sim$ 35 days. For the second group, the feature appears after 40 days, is closer to $H\alpha$ (between 6250 and 6450 \AA), and can last until $\sim$120 days. \citet{Gutierrez_2017} found that Cachito features appearing before 40 days after explosion are due to Si II 6355\AA~for 60$\%$ of the cases and for the rest, it is likely due to high velocity (HV) H$\alpha$. For the SNe that have the Cachito feature 40 days after explosion, \citet{Chugai_2007} has shown the feature is due to HV H absorption.
 SN~2023axu falls in the latter category where the feature develops after 40 days as seen in Fig.~\ref{fig:spec_all}. For this category of SNe, \citet{Gutierrez_2017} found that a similar shallow absorption feature is seen on the blue side of $H\beta$ which we find in our spectrum as well. The time-series evolution of this feature is shown in Fig.~\ref{fig:cachito}.

\begin{figure}
\includegraphics[width=\columnwidth]{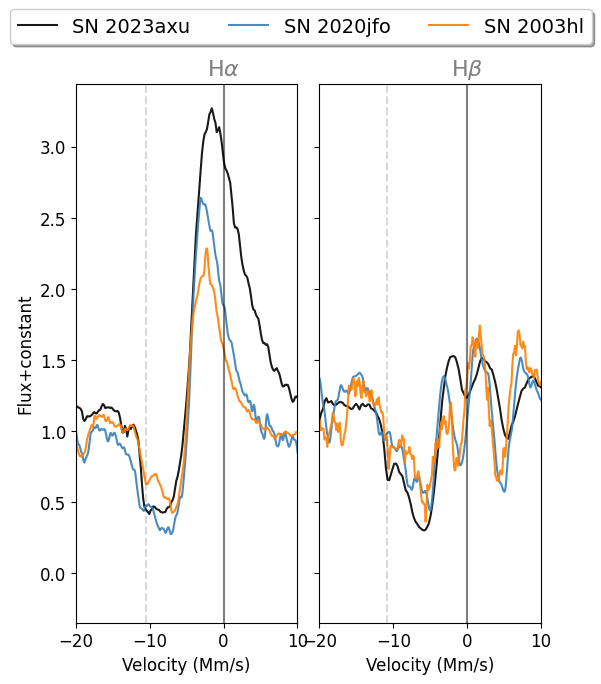}
\caption{Comparison of Cachito feature in SN~2023axu for H$\alpha$ (left) and H$\beta$ (right) shown by the dashed vertical line with SN~2020jfo and SN~2003hl. The epoch of observation are +57.2, +55, and +63 days for SN~2023axu, SN~2020jfo, and SN~2003hl respectively. The velocity is calculated with respect to H$\alpha$ 6562.819 \AA~and H$\beta$ 4861 \AA~shown in a gray solid line. The overall feature of SN~2023axu is similar to SN~2020jfo and SN~2003hl.
\label{fig:cachito_sncomp}}
\end{figure}
Though the SN~2023axu Cachito feature develops after 40 days, and this is usually attributed to a HV H feature, we explored the possibility of a Si II 6355 \AA~ \citep{Valenti_2014} or Ba II 6497 \AA origin. If the feature is arising due to Si II 6355 \AA~then the line velocity inferred is closer to a few hundred km s$^{-1}$. On the other hand, if we assume the Ba II 6497 \AA~line is producing the feature then we infer the velocity to be $\sim$ 7500 km s$^{-1}$. Both of these velocities are different from the velocity inferred from the Fe II 5169 \AA~line ($\sim$ 4000 km s$^{-1}$). In addition, we also find shallow absorption features on the blue side of H$\beta$, hence we disfavour this possibility. 

Finally, we checked if HV H, which originates when X-rays from the SN shock ionize and excite the outer shocked ejecta \citep{Chugai_2007}, can explain this feature. We find this feature in SN~2023axu 40 days after explosion and there is a similar feature on the blue side of $H\beta$ at a similar velocity ($\sim 10500$ $\rm km  s^{-1}$) as the $H\alpha$ case i.e. $\sim$ 10000 $\rm km\ s^{-1}$. This velocity is comparable to the velocity we derived for $H\alpha$ at +28 days after the explosion. \citet{Gutierrez_2017} found 43 out of 122 SNe II show the Cachito later than 40 days in their sample and 63\% of them also have a counterpart in H$\beta$. They find HV H explains the feature the best. In addition, \citet{Kilpatrick_2023} and \citet{Teja_2022} found a similar feature for another Type II SN~2020jfo and \citet{Teja_2022} attributed to HV H as well. Interestingly, SN~2020jfo also contains a broad ledge feature at the first epoch of observation (+3 days).  \citet{Teja_2022} attribute this feature to high ionization of the ejecta and the nearby CSM. In Fig.~\ref{fig:cachito_sncomp}, we find similarities in the H$\alpha$ absorption shape of SN~2023axu to that of SN~2020jfo and SN~2003hl \citep{Gutierrez_2017}. While the absorption feature in H$\beta$ is not seen for SN~2020jfo,  it is present in the spectra of SN~2003hl and is similar to that seen in SN~2023axu. Hence, we favor the HV H scenario for the Cachito feature seen in SN~2023axu.

\section{Discussions \& Conclusion} \label{sec:conclusions}
In this paper, we present the photometric and spectroscopic data analysis of SN~2023axu. The supernova was discovered by DLT40 within 24 hours of the explosion in NGC 2283 at a distance of 13.68 Mpc and prompt spectroscopic follow-up using PyMMT and other instruments enabled us to obtain a spectrum at +1.1 days after the explosion. 
The SN is a typical Type II with an absolute peak $V$-band magnitude $-$16.53 and a plateau phase lasting 101 days. Our observations extended into the radioactive tail phase, allowing us to calculate the nickel mass by comparing the pseudo-bolometric lightcurve of SN~2023axu with SN~1987A. We found the nickel mass to be 0.029 $\pm$ 0.01 M$\odot$. This value is within a typical range for SNe II of 0.003 to 0.17 M$\odot$ \citep{Valenti_2016}. 

We also performed MCMC shock cooling fitting to the early light curve of SN~2023axu following the prescription by \citet{Morang_2023} and implemented by \citet{Hosseinzadeh_2023_23ixf}. We find the model to converge, however, the model under-predicts the signal for the UV data and the steep rise in the $r,i$ filters. The observed excess flux could be attributed to CSM interaction. From our light curve analysis with shock-cooling models we find the progenitor radius to be  417 $\pm$ 28 R$\odot$ implying a RSG progenitor \citep{Levesque_2017}. RSGs are known for mass loss during their lifetime. There is increasing evidence of the presence of CSM in many SNe II via 
the presence of a steep rise in their light curves that can not be explained by shock-cooling models that assume a lack of CSM. \citet{Morozova_2017,Morozova_2018} found the presence of dense CSM in their numerical setup can accurately model the observed early steep rise in the light curve. 

We also present high-cadence optical spectra starting at +1.1 to +112.3 days after the explosion epoch. In general, the spectral evolution is typical for SNe II with two notable features: 1) The first two epochs of spectra show the ledge feature at 4400-4800 \AA. 
The ledge feature has been seen in other type II SNe \citep[e.g.][]{Andrews_2019,Hosseinzadeh_2022_21yja,Pearson_2023,Bostroem_2023} and has been attributed to the presence of CSM. We also compared this behavior to models by \citet{Dessart_2017} and found the \texttt{r1w1} model with a low-density CSM and $\dot{M} = 10^{-6} M_{\odot}$ yr$^{-1}$ behaves closest to the observed early spectra of SN~2023axu. 2) At epochs $>$40 days, we see shallow absorption features on the blue side of H$\alpha$ and H$\beta$ and we interpret this as HV H.  \citet{Chugai_2007} proposed this feature to be the result of the interaction between the SN ejecta and the RSG wind, thus implying the presence of CSM interaction for the case of SN~2023axu.  Both spectroscopic and photometric features point to the most likely scenario of an explosion of an RSG surrounded by low-density CSM for SN~2023axu. The steep rise in the early light-curve and ledge features in early spectra probe the mass loss from RSG in the final phases before the explosion whereas the Cachito feature probes the CSM produced by RSG mass loss earlier in its evolution. This work adds to the growing evidence of the presence of CSM around the progenitors of SNe II.

The combination of high-cadence multi-wavelength photometric and rapid spectroscopic data helped us constrain the properties of SN~2023axu, including the likely presence of CSM. These results show the need for high-cadence photometric observations of SNe II along with infrastructure for rapid spectroscopic follow-up. Implementing infrastructure like {\sc PyMMT} is critical to improving our understanding of the final stages of RSG evolution to SN.


\section{Acknowledgments}
Time-domain research by the University of Arizona team, M.S. and D.J.S.\ is supported by NSF grants AST-1821987, 1813466, 1908972, 2108032, and 2308181, and by the Heising-Simons Foundation under grant \#2020-1864. K.A.B. is supported by an LSSTC Catalyst Fellowship; this publication was thus made possible through the support of Grant 62192 from the John Templeton Foundation to LSSTC. The opinions expressed in this publication are those of the authors and do not necessarily reflect the views of LSSTC or the John Templeton Foundation. 
Research by Y.D., S.V., N.M.R, E.H., and D.M. is supported by NSF grant AST-2008108. 
 This research has made use of the NASA Astrophysics Data System (ADS) Bibliographic Services, and the NASA/IPAC Infrared Science Archive (IRSA), which is funded by the National Aeronautics and Space Administration and operated by the California Institute of Technology.  This research made use of Photutils, an Astropy package for detection and photometry of astronomical sources (\cite{Bradley_2019}). This work made use of data supplied by the UK Swift Science Data Centre at the University of Leicester.
Observations reported here were obtained at the MMT Observatory, a joint facility of the University of Arizona and the Smithsonian Institution.  
Based in part on observations obtained at the Southern Astrophysical Research (SOAR) telescope, which is a joint project of the  Minist\'{e}rio da Ci\^{e}ncia, Tecnologia e Inova\c{c}\~{o}es (MCTI/LNA) do Brasil, the US National Science Foundation's NOIRLab, the University of North Carolina at Chapel Hill (UNC), and Michigan State University (MSU). 
This research has made use of the CfA Supernova Archive, which is funded in part by the National Science Foundation through grant AST 0907903. 
This work makes use of data taken with the Las Cumbres Observatory global telescope network. The LCO group is supported by NSF grants 1911225 and 1911151. 
JEA and CEMV are supported by the international Gemini Observatory, a program of NSF's NOIRLab, which is managed by the Association of Universities for Research in Astronomy (AURA) under a cooperative agreement with the National Science Foundation, on behalf of the Gemini partnership of Argentina, Brazil, Canada, Chile, the Republic of Korea, and the United States of America. JAC-B acknowledges support from FONDECYT Regular N 1220083.
The SALT data presented here were obtained via Rutgers University program 2022-1-MLT-004 (PI: SWJ). LAK acknowledges support by NASA FINESST fellowship 80NSSC22K1599.


%

\vspace{5mm}
\facilities{ADS,  MMT (Binospec),  Las Cumbres Observatory,  SOAR (GHTS), SALT (RSS), NED, WISeREP, IRSA }

\software{ astropy \citep{astropy:2013,astropy:2018}, Photutils \citep{Bradley_2019}, Binospec IDL \citep{BinoIDL}, Panacea, BANZAI \citep{Banzai}, Light Curve Fitting \citep{lightcurvefitting}, MatPLOTLIB \citep{mpl}, NumPy \citep{numpy}, Scipy \citep{scipy}, IRAF \citep{iraf1,iraf2}, PySALT \citep{PySALT}}



\appendix

\section{PyMMT} \label{sec:appendix}
We are in the era of focused and general-purpose time domain surveys which have led to an explosion in the number of transients discovered on a nightly basis, however, most of them do not receive any spectroscopic follow-up or classification (see the extensive discussion in \citealt{Kulkarni_2020}). For various science cases such as young supernovae, kilonovae candidates etc. there is an acute need for nearly real-time spectroscopic observations. To decrease the gap between discovery and spectroscopic response various robotic spectrographs such as a) twin robotic FLOYDS spectrographs on the Faulkes Telescope North \& South \citep{Brown_2013}; b) Spectrograph for the Rapid Acquisition of Transients on the Liverpool Telescope (SPRAT; \citet{Piascik_2014}); and c) the Spectral Energy Distribution Machine on the Palomar 60-in telescope (SEDM; \citet{Blagorodnova_2018} have been deployed on 2-m class telescopes capable of rapid target of opportunity observations. When the sources are fainter or we need a better signal-to-noise ratio, a larger aperture size is needed. Several large facilities, such as the South African Large Telescope (SALT), Keck Observatory, the two Gemini telescopes \citep{Roth_2009}, and SOAR have rapid target of opportunity (ToO) programs in place for this purpose. 

The 6.5-meter MMT \footnote{http://www.mmto.org} can potentially play an important role in the rapid follow-up of transients. The observatory has been mostly operating in a queue mode in observing blocks for the past several years for three spectrographs: Binospec; an optical spectrograph \citep{Fabricant_2019}, MMIRS; a NIR spectrograph \citep{McLeod_2012}, and Hectospec; multi-object optical spectrograph \citep{Fabricant_2005}.

\begin{figure*}
\includegraphics[width=\textwidth]{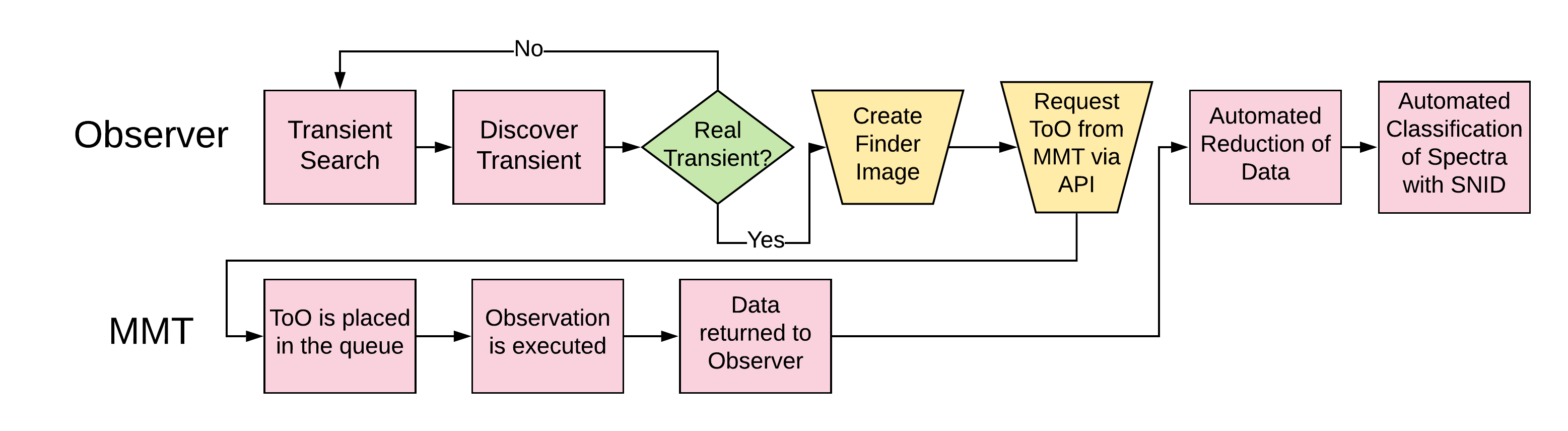}
\caption{Flowchart for transient identification, MMT+Binospec+MMIRS triggering and initial analysis for our program.
\label{fig:flowchart}}
\end{figure*}

Even though the potential for real-time spectroscopic follow-up with MMT is great, the infrastructure is lagging. Currently, in order to request an urgent observation, the observer needs to navigate multiple web pages and submit a finding chart before submitting for the queue. In addition, the observed data are not available as the observations are completed, which can be crucial for certain science cases. All of these extra steps slow down the activation of a ToO request. Hence, we have developed PyMMT \citep{samuel_wyatt_2023_8322354}, a Python package that communicates with the MMT scheduling software's application programming interface \citep[API;][]{Gibson_2018}, allowing direct, seamless injection of new targets into the observation queue for rapid follow-up.  Currently, PyMMT has the capability to trigger Binospec 
and MMIRS. The work flow of PyMMT is shown in Fig.~\ref{fig:flowchart}. 

PyMMT communicates with four endpoints of the MMT API. Each is encapsulated in a Python class. Note that, for all classes (except \verb|Instruments|; Appendix~\ref{sec:instruments}), the user must provide an API token, either by setting the \verb|token| keyword at initialization or by setting the \verb|MMT_API_TOKEN| environment variable. We briefly describe the functionality of each class below. For full documentation, see the \texttt{README.md} file in GitHub.\footnote{\url{https://github.com/SAGUARO-MMA/PyMMT}}

\subsection{Requesting Observations}\label{sec:target}
The \verb|Target| class allows the user to submit observation requests to the queue and query their status.\footnote{In the language of the MMT API, a ``target'' is not a unique pair of coordinates, but rather a unique request for an observation of a pair of coordinates. One might submit multiple ``targets'' corresponding to the same SN, for example, to obtain a series of observations at different times or with different configurations.} It contains validation routines for requested configurations on the two supported instruments, Binospec and MMIRS. To initialize a new \verb|Target| instance, the user first assembles a payload containing the name, coordinates, and brightness of the target, along with instrument-specific configuration parameters. An example MMIRS spectroscopy payload is shown below:
\begin{verbatim}
payload = {
    # target details
    'objectid': 'SN2023axu',
    'ra': '06:45:55.320',
    'dec': '-18:13:53.50',
    'epoch': 'J2000',
    'magnitude': 21.,

    # scheduling details 
    'notes': 'Demo observation request.',
    'priority': 3,
    'targetofopportunity': 0,
    'visits': 1,

    # instrument and mode specification
    'instrumentid': 15,
    'observationtype': 'longslit',

    # instrument-specific configuration
    'dithersize': '5',
    'exposuretime': 450.,
    'filter': 'zJ',
    'gain': 'low',
    'grism': 'J',
    'maskid': 111,
    'numberexposures': 3,
    'readtab': 'ramp_4.426',
    'slitwidth': '1pixel',
    'slitwidthproperty': 'long',   
}
\end{verbatim}
If a target has already been submitted, its details can be retrieved from the MMT API by providing only its ID number, e.g., \verb|payload = {'targetid': 14294}|.

The \verb|Target| instance is then initialized with
\verb|target = pymmt.Target(payload=payload)|. 
If a target ID is provided in the payload, details of the already submitted target will be retrieved from the MMT API immediately on initialization, or they can be retrieved manually with \verb|target.get()|.
Validation is also run immediately on initialization, or it can be run manually with \verb|target.validate()|. If any of the provided parameters are invalid, an error message will be printed to the screen.
At any point, the parameters can be printed to the screen with \verb|target.dump()|.
The target can be submitted to the MMT API with \verb|target.post()|.
A finder chart, which is required for spectroscopic observations, can be uploaded with, e.g., \verb|target.upload_finder('/path/to/finder.png')|.
Updated target parameters can be sent to the API with, e.g., \verb|target.update(magnitude=22.)|.
A target can be deleted (i.e., the observation request can be cancelled) with \verb|target.delete()|.
Finally, data for a target can be downloaded with \verb|target.download_exposures()| (see also Appendix~\ref{sec:datalist} below). The data will appear in the \verb|data| subdirectory of the current working directory, e.g., \verb|./data/SN2023axu/D2023.0403/FITS Image/|.

\subsection{Viewing the Schedule}\label{sec:instruments}
The \verb|Instruments| class retrieves and parses the MMT schedule, so users can see which instrument is available on a given night. Because the schedule is public, no API token is required. After initializing an \verb|Instruments| instance with \verb|insts = pymmt.Instruments()|, the schedule can be queried either by instrument or by date. A call to \verb|insts.get_instruments(instrumentid=16)| will return a list of schedule entries indicating when Binospec is on the telescope, as well as print them to the screen. Each entry includes the instrument ID (\verb|'instrumentid'|), program name (\verb|'name'|), and start and end times (\verb|'start'| and \verb|'end'|), given as Python \verb|datetime| instances. A call to \verb|insts.get_instruments(date=datetime(2023, 11, 7))| will return and print the schedule entry for the given date and time. A call to \verb|insts.get_instruments()| with no arguments returns and prints the currently active program.

\subsection{Retrieving Data}\label{sec:datalist}
After observations are obtained, users can view their metadata using the \verb|Datalist| class. After initializing a \verb|Datalist| instance with \verb|datalist = pymmt.Datalist()|, a list of raw data products for a given target can be obtained with, e.g., \verb|datalist.get(targetid=14294)|. Reduced data products can be substituted by giving the keyword argument \verb|data_type='reduced'|. Metadata for these products is stored in the \verb|datalist.data| attribute. When initializing an already observed \verb|Target|, metadata is automatically stored in the \verb|target.datalist.data| attribute.

Users can download their data using the \verb|Image| class. After initializing an \verb|Image| instance with \verb|im = pymmt.Image()|, a single data product can be downloaded to a local file \verb|'data.fits'| using \verb|im.get(datafileid=fid, filepath='data.fits')|, where \verb|fid| is the ID number of the data product from, e.g., \verb|datalist.data[0]['id']|. We recommend using the higher-level \verb|target.download_exposures()| (Appendix~\ref{sec:target}) for downloading raw data of a target in bulk.

\bibliography{SN2023axu.bib}{}
\bibliographystyle{aasjournal}



\end{document}

%% file: affiliation.tex
\newcommand{\LCO}{\affiliation{Las Cumbres Observatory, 6740 Cortona Drive, Suite 102, Goleta, CA 93117-5575, USA}}
\newcommand{\UCSB}{\affiliation{Department of Physics, University of California, Santa Barbara, CA 93106-9530, USA}}
\newcommand{\KITP}{\affiliation{Kavli Institute for Theoretical Physics, University of California, Santa Barbara, CA 93106-4030, USA}}
\newcommand{\UCD}{\affiliation{Department of Physics and Astronomy, University of California, Davis, 1 Shields Avenue, Davis, CA 95616-5270, USA}}
\newcommand{\WIS}{\affiliation{Department of Particle Physics and Astrophysics, Weizmann Institute of Science, 76100 Rehovot, Israel}}
\newcommand{\OKC}{\affiliation{Oskar Klein Centre, Department of Astronomy, Stockholm University, Albanova University Centre, SE-106 91 Stockholm, Sweden}}
\newcommand{\OAPD}{\affiliation{INAF-Osservatorio Astronomico di Padova, Vicolo dell'Osservatorio 5, I-35122 Padova, Italy}}
\newcommand{\Caltech}{\affiliation{Cahill Center for Astronomy and Astrophysics, California Institute of Technology, Mail Code 249-17, Pasadena, CA 91125, USA}}
\newcommand{\GSFC}{\affiliation{Astrophysics Science Division, NASA Goddard Space Flight Center, Mail Code 661, Greenbelt, MD 20771, USA}}
\newcommand{\UMD}{\affiliation{Joint Space-Science Institute, University of Maryland, College Park, MD 20742, USA}}
\newcommand{\UCB}{\affiliation{Department of Astronomy, University of California, Berkeley, CA 94720-3411, USA}}
\newcommand{\TTU}{\affiliation{Department of Physics, Texas Tech University, Box 41051, Lubbock, TX 79409-1051, USA}}
\newcommand{\STScI}{\affiliation{Space Telescope Science Institute, 3700 San Martin Drive, Baltimore, MD 21218-2410, USA}}
\newcommand{\UT}{\affiliation{University of Texas at Austin, 1 University Station C1400, Austin, TX 78712-0259, USA}}
\newcommand{\IoA}{\affiliation{Institute of Astronomy, University of Cambridge, Madingley Road, Cambridge CB3 0HA, UK}}
\newcommand{\QUB}{\affiliation{Astrophysics Research Centre, School of Mathematics and Physics, Queen's University Belfast, Belfast BT7 1NN, UK}}
\newcommand{\IPAC}{\affiliation{Spitzer Science Center, California Institute of Technology, Pasadena, CA 91125, USA}}
\newcommand{\JPL}{\affiliation{Jet Propulsion Laboratory, California Institute of Technology, 4800 Oak Grove Dr, Pasadena, CA 91109, USA}}
\newcommand{\Southampton}{\affiliation{Department of Physics and Astronomy, University of Southampton, Southampton SO17 1BJ, UK}}
\newcommand{\LANL}{\affiliation{Space and Remote Sensing, MS B244, Los Alamos National Laboratory, Los Alamos, NM 87545, USA}}
\newcommand{\Tsinghua}{\affiliation{Physics Department and Tsinghua Center for Astrophysics, Tsinghua University, Beijing, 100084, People's Republic of China}}
\newcommand{\NAOC}{\affiliation{National Astronomical Observatory of China, Chinese Academy of Sciences, Beijing, 100012, People's Republic of China}}
\newcommand{\Itagaki}{\affiliation{Itagaki Astronomical Observatory, Yamagata 990-2492, Japan}}
\newcommand{\Einstein}{\altaffiliation{Einstein Fellow}}
\newcommand{\Hubble}{\altaffiliation{Hubble Fellow}}
\newcommand{\CfA}{\affiliation{Center for Astrophysics \textbar{} Harvard \& Smithsonian, 60 Garden Street, Cambridge, MA 02138-1516, USA}}
\newcommand{\UA}{\affiliation{Steward Observatory, University of Arizona, 933 North Cherry Avenue, Tucson, AZ 85721-0065, USA}}
\newcommand{\MPIA}{\affiliation{Max-Planck-Institut f\"ur Astrophysik, Karl-Schwarzschild-Stra\ss{}e 1, D-85748 Garching, Germany}}
\newcommand{\DSFP}{\altaffiliation{LSSTC Data Science Fellow}}
\newcommand{\HCO}{\affiliation{Harvard College Observatory, 60 Garden Street, Cambridge, MA 02138-1516, USA}}
\newcommand{\Carnegie}{\affiliation{Observatories of the Carnegie Institute for Science, 813 Santa Barbara Street, Pasadena, CA 91101-1232, USA}}
\newcommand{\TAU}{\affiliation{School of Physics and Astronomy, Tel Aviv University, Tel Aviv 69978, Israel}}
\newcommand{\Edinburgh}{\affiliation{Institute for Astronomy, University of Edinburgh, Royal Observatory, Blackford Hill EH9 3HJ, UK}}
\newcommand{\Birmingham}{\affiliation{Birmingham Institute for Gravitational Wave Astronomy and School of Physics and Astronomy, University of Birmingham, Birmingham B15 2TT, UK}}
\newcommand{\Bath}{\affiliation{Department of Physics, University of Bath, Claverton Down, Bath BA2 7AY, UK}}
\newcommand{\CTIO}{\affiliation{Cerro Tololo Inter-American Observatory, National Optical Astronomy Observatory, Casilla 603, La Serena, Chile}}
\newcommand{\Potsdam}{\affiliation{Institut f\"ur Physik und Astronomie, Universit\"at Potsdam, Haus 28, Karl-Liebknecht-Str. 24/25, D-14476 Potsdam-Golm, Germany}}
\newcommand{\INPE}{\affiliation{Instituto Nacional de Pesquisas Espaciais, Avenida dos Astronautas 1758, 12227-010, S\~ao Jos\'e dos Campos -- SP, Brazil}}
\newcommand{\UNC}{\affiliation{Department of Physics and Astronomy, University of North Carolina, 120 East Cameron Avenue, Chapel Hill, NC 27599, USA}}
\newcommand{\Ohio}{\affiliation{Astrophysical Institute, Department of Physics and Astronomy, 251B Clippinger Lab, Ohio University, Athens, OH 45701-2942, USA}}
\newcommand{\AAS}{\affiliation{American Astronomical Society, 1667 K~Street NW, Suite 800, Washington, DC 20006-1681, USA}}
\newcommand{\MMT}{\affiliation{MMT and Steward Observatories, University of Arizona, 933 North Cherry Avenue, Tucson, AZ 85721-0065, USA}}
\newcommand{\Geneva}{\affiliation{ISDC, Department of Astronomy, University of Geneva, Chemin d'\'Ecogia, 16 CH-1290 Versoix, Switzerland}}
\newcommand{\IUCAA}{\affiliation{Inter-University Center for Astronomy and Astrophysics, Post Bag 4, Ganeshkhind, Pune, Maharashtra 411007, India}}
\newcommand{\CMU}{\affiliation{Department of Physics, Carnegie Mellon University, 5000 Forbes Avenue, Pittsburgh, PA 15213-3815, USA}}
\newcommand{\NAOJ}{\affiliation{Division of Science, National Astronomical Observatory of Japan, 2-21-1 Osawa, Mitaka, Tokyo 181-8588, Japan}}
\newcommand{\IfA}{\affiliation{Institute for Astronomy, University of Hawai`i, 2680 Woodlawn Drive, Honolulu, HI 96822-1839, USA}}
\newcommand{\UCSC}{\affiliation{Department of Astronomy and Astrophysics, University of California, Santa Cruz, CA 95064-1077, USA}}
\newcommand{\Purdue}{\affiliation{Department of Physics and Astronomy, Purdue University, 525 Northwestern Avenue, West Lafayette, IN 47907-2036, USA}}
\newcommand{\Princeton}{\affiliation{Department of Astrophysical Sciences, Princeton University, 4 Ivy Lane, Princeton, NJ 08540-7219, USA}}
\newcommand{\Moore}{\affiliation{Gordon and Betty Moore Foundation, 1661 Page Mill Road, Palo Alto, CA 94304-1209, USA}}
\newcommand{\Durham}{\affiliation{Department of Physics, Durham University, South Road, Durham, DH1 3LE, UK}}
\newcommand{\JHU}{\affiliation{Department of Physics and Astronomy, The Johns Hopkins University, 3400 North Charles Street, Baltimore, MD 21218, USA}}
\newcommand{\Toronto}{\affiliation{David A.\ Dunlap Department of Astronomy and Astrophysics, University of Toronto,\\ 50 St.\ George Street, Toronto, Ontario, M5S 3H4 Canada}}
\newcommand{\Duke}{\affiliation{Department of Physics, Duke University, Campus Box 90305, Durham, NC 27708, USA}}
\newcommand{\NCU}{\affiliation{Graduate Institute of Astronomy, National Central University, 300 Jhongda Road, 32001 Jhongli, Taiwan}}
\newcommand{\Columbia}{\affiliation{Department of Physics and Columbia Astrophysics Laboratory, Columbia University, Pupin Hall, New York, NY 10027, USA}}
\newcommand{\Flatiron}{\affiliation{Center for Computational Astrophysics, Flatiron Institute, 162 5th Avenue, New York, NY 10010-5902, USA}}
\newcommand{\CIERA}{\affiliation{Center for Interdisciplinary Exploration and Research in Astrophysics and Department of Physics and Astronomy, \\Northwestern University, 1800 Sherman Avenue, 8th Floor, Evanston, IL 60201, USA}}
\newcommand{\GeminiNorth}{\affiliation{Gemini Observatory, 670 North A`ohoku Place, Hilo, HI 96720-2700, USA}}
\newcommand{\Keck}{\affiliation{W.~M.~Keck Observatory, 65-1120 M\=amalahoa Highway, Kamuela, HI 96743-8431, USA}}
\newcommand{\UW}{\affiliation{Department of Astronomy, University of Washington, 3910 15th Avenue NE, Seattle, WA 98195-0002, USA}}
\newcommand{\catalyst}{\altaffiliation{LSSTC Catalyst Fellow}}
\newcommand{\USask}{\affiliation{Department of Physics \& Engineering Physics, University of Saskatchewan, 116 Science Place, Saskatoon, SK S7N 5E2, Canada}}
\newcommand{\Thacher}{\affiliation{Thacher School, 5025 Thacher Road, Ojai, CA 93023-8304, USA}}
\newcommand{\Rutgers}{\affiliation{Department of Physics and Astronomy, Rutgers, the State University of New Jersey,\\136 Frelinghuysen Road, Piscataway, NJ 08854-8019, USA}}
\newcommand{\FSU}{\affiliation{Department of Physics, Florida State University, 77 Chieftan Way, Tallahassee, FL 32306-4350, USA}}
\newcommand{\Melbourne}{\affiliation{School of Physics, The University of Melbourne, Parkville, VIC 3010, Australia}}
\newcommand{\ASTROthreeD}{\affiliation{ARC Centre of Excellence for All Sky Astrophysics in 3 Dimensions (ASTRO 3D)}}
\newcommand{\Stromlo}{\affiliation{Mt.\ Stromlo Observatory, The Research School of Astronomy and Astrophysics, Australian National University, ACT 2601, Australia}}
\newcommand{\NCPAS}{\affiliation{National Centre for the Public Awareness of Science, Australian National University, Canberra, ACT 2611, Australia}}
\newcommand{\TAMU}{\affiliation{Department of Physics and Astronomy, Texas A\&M University, 4242 TAMU, College Station, TX 77843, USA}}
\newcommand{\Mitchell}{\affiliation{George P.\ and Cynthia Woods Mitchell Institute for Fundamental Physics \& Astronomy, College Station, TX 77843, USA}}
\newcommand{\ESO}{\affiliation{European Southern Observatory, Alonso de C\'ordova 3107, Casilla 19, Santiago, Chile}}
\newcommand{\ICE}{\affiliation{Institute of Space Sciences (ICE, CSIC), Campus UAB, Carrer
de Can Magrans, s/n, E-08193 Barcelona, Spain}}
\newcommand{\IEEC}{\affiliation{Institut d'Estudis Espacials de Catalunya, Gran Capit\`a, 2-4, Edifici Nexus, Desp.\ 201, E-08034 Barcelona, Spain}}
\newcommand{\Warwick}{\affiliation{Department of Physics, University of Warwick, Gibbet Hill Road, Coventry CV4 7AL, UK}}
\newcommand{\Macquarie}{\affiliation{School of Mathematical and Physical Sciences, Macquarie University, NSW 2109, Australia}}
\newcommand{\AAARC}{\affiliation{Astronomy, Astrophysics and Astrophotonics Research Centre, Macquarie University, Sydney, NSW 2109, Australia}}
\newcommand{\Capodimonte}{\affiliation{INAF - Capodimonte Astronomical Observatory, Salita Moiariello 16, I-80131 Napoli, Italy}}
\newcommand{\INFNNapoli}{\affiliation{INFN - Napoli, Strada Comunale Cinthia, I-80126 Napoli, Italy}}
\newcommand{\ICRANet}{\affiliation{ICRANet, Piazza della Repubblica 10, I-65122 Pescara, Italy}}
\newcommand{\MSU}{\affiliation{Center for Data Intensive and Time Domain Astronomy, Department of Physics and Astronomy,\\Michigan State University, East Lansing, MI 48824, USA}}
\newcommand{\SETI}{\affiliation{SETI Institute,
339 Bernardo Ave, Suite 200, Mountain View, CA 94043, USA}}
\newcommand{\IAIFI}{\affiliation{The NSF AI Institute for Artificial Intelligence and Fundamental Interactions}}
\newcommand{\ANUC}{\affiliation{Department of Astronomy, AlbaNova University Center, Stockholm University, SE-10691 Stockholm, Sweden}}

\newcommand{\Konkoly}{\affiliation{Konkoly Observatory,  CSFK, Konkoly-Thege M. \'ut 15-17, Budapest, 1121, Hungary}}
\newcommand{\ELTE}{\affiliation{ELTE E\"otv\"os Lor\'and University, Institute of Physics, P\'azm\'any P\'eter s\'et\'any 1/A, Budapest, 1117 Hungary}}
\newcommand{\SZTE}{\affiliation{Department of Experimental Physics, University of Szeged, D\'om t\'er 9, Szeged, 6720, Hungary}}
\newcommand{\IdAlta}{\affiliation{Instituto de Alta Investigaci\'on, Sede Esmeralda, Universidad de Tarapac\'a, Av. Luis Emilio Recabarren 2477, Iquique, Chile}}
\newcommand{\Kavli}{\affiliation{Kavli Institute for Cosmological Physics, University of Chicago, Chicago, IL 60637, USA}}
\newcommand{\UofChicago}{\affiliation{Department of Astronomy and Astrophysics, University of Chicago, Chicago, IL 60637, USA}}
\newcommand{\Fermi}{\affiliation{Fermi National Accelerator Laboratory, P.O.\ Box 500, Batavia, IL 60510, USA}}
\newcommand{\Dartmouth}{\affiliation{Department of Physics and Astronomy, Dartmouth College, Hanover, NH 03755, USA}}
\newcommand{\Surrey}{\affiliation{Department of Physics, University of Surrey, Guildford GU2 7XH, UK}}
\newcommand{\NU}{\affiliation{Center for Interdisciplinary Exploration and Research in Astrophysics (CIERA) and Department of Physics and Astronomy, Northwestern University, Evanston, IL 60208, USA}}

\newcommand{\itagaki}{\affiliation{Itagaki Astronomical Observatory, Yamagata 990-2492, Japan}}

%% file: author.tex
\author[0000-0002-4022-1874]{Manisha Shrestha}
\UA

\author[0000-0002-0744-0047]{Jeniveve Pearson}
\UA
\author[0000-0003-2732-4956]{Samuel Wyatt}
\UW

\author[0000-0003-4102-380X]{David J. Sand}
\UA
\author[0000-0002-0832-2974]{Griffin Hosseinzadeh}
\UA

\author[0000-0002-4924-444X]{K. Azalee Bostroem}
\catalyst\UA

\author[0000-0003-0123-0062]{Jennifer E. Andrews}
\GeminiNorth

\author[0000-0002-7937-6371]{Yize Dong \begin{CJK*}{UTF8}{gbsn}(董一泽)\end{CJK*}}
\UCD
\author[0000-0003-2744-4755]{Emily Hoang}
\UCD

\author[0000-0003-0549-3281]{Daryl Janzen}
\USask
\author[0000-0001-5754-4007]{Jacob E. Jencson}
\JHU

\author[0000-0001-9589-3793]{M.~J. Lundquist}

\Keck
\author{Darshana Mehta}
\UCD

\author[0000-0002-7015-3446]{Nicol\'as Meza Retamal}
\UCD

\author[0000-0001-8818-0795]{Stefano Valenti}
\UCD

\author[0000-0002-9267-6213]{Jillian C. Rastinejad}
\NU

\author{Phil Daly}
\UA
\author{Dallan Porter}
\MMT
\author{Joannah Hinz}
\MMT
\author{Skyler Self}
\MMT
\author[0000-0001-6065-7483]{Benjamin Weiner}
\MMT
\author[0000-0002-3452-0560]{Grant G. Williams}
\MMT
\UA

\author[0000-0002-1125-9187]{Daichi Hiramatsu}
\CfA
\IAIFI

\author[0000-0003-4253-656X]{D.\ Andrew Howell}
\LCO\UCSB
\author[0000-0001-5807-7893]{Curtis McCully}
\LCO
\UCSB

\author[0000-0003-0209-9246]{Estefania Padilla Gonzalez}
\LCO
\UCSB
\author[0000-0002-7472-1279]{Craig Pellegrino}
\LCO 
\UCSB
\author[0000-0003-0794-5982]{Giacomo Terreran}
\LCO 
\UCSB

\author{Megan Newsome}
\LCO 
\UCSB

\author{Joseph Farah}
\LCO 
\UCSB

\author{Koichi Itagaki}
\itagaki

\author[0000-0001-8738-6011]{Saurabh W.\ Jha}
\Rutgers

\author[0000-0003-3108-1328]{Lindsey Kwok}
\Rutgers

\author[0000-0001-5510-2424]{Nathan Smith}
\UA

\author[0009-0002-5096-1689]{Michaela Schwab}
\Rutgers

\author[0000-0003-3643-839X]{Jeonghee Rho}
\SETI
\author[0000-0002-6535-8500]{Yi Yang}
\UCB